\documentclass[a4paper,11pt]{article}
\pdfoutput=1
\usepackage{jheppub,fancyhdr,graphicx,epstopdf,color,amsmath,cases,slashed,hyperref,multirow}
\usepackage{cancel}
\usepackage{float}
\usepackage{enumerate}
\usepackage{epsfig}
\usepackage{amsmath} 
\usepackage{amssymb} 
\usepackage{wasysym} 
\usepackage{mathrsfs} 
\usepackage{graphicx} 
\usepackage{amsfonts} 
\usepackage{amsbsy} 
\usepackage{amscd}
\usepackage{slashed} 
\usepackage{tikz}
\usepackage{multirow} 
\usepackage{color}
\definecolor{myred}{rgb}{0.6,0,0} 
\definecolor{myblue}{rgb}{0,0.2,0.4}
\definecolor{mygreen}{rgb}{0,0.9,0.1}
\definecolor{hc}{rgb}{.9,0.1,0.7}
\definecolor{hcout}{rgb}{.9,0.7,0.9}
\definecolor{Orange}{rgb}{1.,0.65,0.}

\makeatletter

\newcommand{\fmslash}[2][0mu]{%
  \mathchoice
    {\fmsl@sh\displaystyle{#1}{#2}}%
    {\fmsl@sh\textstyle{#1}{#2}}%
    {\fmsl@sh\scriptstyle{#1}{#2}}%
    {\fmsl@sh\scriptscriptstyle{#1}{#2}}}
\newcommand{\fmsl@sh}[3]{%
  \m@th\ooalign{$\hfil#1\mkern#2/\hfil$\crcr$#1#3$}}
\makeatother

\newcommand{\lsim}{{\;\raise0.3ex\hbox{$<$\kern-0.75em\raise-1.1ex\hbox{$\sim$}}\;}}
\newcommand{\gsim}{{\;\raise0.3ex\hbox{$>$\kern-0.75em\raise-1.1ex\hbox{$\sim$}}\;}}

\usepackage{array}
\newcolumntype{C}[1]{>{\centering\arraybackslash$}p{#1}<{$}}

\usepackage{bm} 
\newcommand{\be}{\begin{equation}}
\newcommand{\ee}{\end{equation}}
\newcommand{\bes}{\begin{equation*}}
\newcommand{\ees}{\end{equation*}}
\newcommand{\bea}{\begin{eqnarray}}
\newcommand{\eea}{\end{eqnarray}}
\newcommand{\beas}{\begin{eqnarray*}}
\newcommand{\eeas}{\end{eqnarray*}}

\newcommand{\ba}{\begin{array}}
\newcommand{\ea}{\end{array}}
\newcommand{\bd}{\begin{displaymath}}
\newcommand{\ed}{\end{displaymath}}
\newcommand{\besub}{\begin{subequations}}
\newcommand{\eesub}{\end{subequations}}

\def\a{\alpha}

\def\g{\gamma}

\def\l{\lambda}

\def\q2 {q^2}

\def\bt{\begin{table}}
\def\et{\end{table}}

\title{A revisit to scalar dark matter with radiative corrections} 

\preprint{LAPTH-069/16, HRI-RECAPP-2016-006, IPPP/18/80, NCTS-PH/1813}

\author[a,b]{Shankha Banerjee}
\author[c,d]{, Nabarun Chakrabarty}

  \affiliation[a]{Universit\'{e} Grenoble Alpes, USMB, CNRS, LAPTh, F-74000 Annecy, France}
  \affiliation[b]{Institute for Particle Physics Phenomenology, Department of Physics, Durham University, Durham DH1 3LE, United Kingdom}
  \affiliation[c]{Regional Centre for Accelerator-based Particle Physics, Harish-Chandra Research Institute, HBNI, Chhatnag Road, Jhusi, Allahabad 211019, India}
  \affiliation[d]{Physics Division, National Center for Theoretical Sciences, Kuang-Fu Road, Hsinchu, Taiwan 30013, R.O.C.}
\emailAdd{shankha.banerjee@durham.ac.uk}
\emailAdd{nchakrabarty@cts.nthu.edu.tw}

\abstract{Extended Higgs sectors have been studied extensively in context of dark matter phenomenology in tandem with other aspects. In this study, we compute radiative corrections to the dark matter-Higgs portal coupling, which is in fact a common feature of all scalar dark matter models irrespective of the hypercharge of the multiplet from which the dark matter candidate emerges. We select the popular inert doublet model (IDM) as a prototype in order to demonstrate the impact of the next-to-leading order corrections, thereby probing the plausibility of extending the allowed parameter space through quantum effects. Given that the tree level portal coupling is a \emph{prima facie} free parameter, the percentage change from loop effects can be large. This modifies the dark matter phenomenology at a quantitative level. It also encourages one to include loop corrections to all other interactions that are deemed relevant in this context.}

\begin{document}

\maketitle
\newpage

\section{Introduction}
\label{sec1}

The successful discovery of the Higgs boson in 2012 by the CMS~\cite{Chatrchyan:2012xdj} and ATLAS~\cite{Aad:2012tfa} collaborations completed the search for the last missing piece in the Standard Model (SM) of particle physics. However, SM is unable to answer certain fundamental observations, \textit{viz.}, the existence of dark matter (DM), massive neutrinos, the excess of baryons over anti-baryons, three generations of leptons etc. Besides, there are certain theoretical issues like the hierarchy problem, which the SM fails to answer. We are thus led to consider physics beyond the standard model (BSM) in order to explain such observations. After the discovery of the Higgs boson, the theoretical and experimental community have spent all their resources in studying the couplings and the $CP$ nature of this discovered boson. The coupling strengths of the Higgs boson to other SM particles conform with their SM expectations within 1$\sigma$. A purely $CP$-odd scenario is also shown to be disfavoured by experiments. The invisible branching ratio of an SM-like Higgs boson is also constrained by experiments and global fits to $\sim 20\%$ at 95\% CL~\cite{Aaboud:2017bja, Aaboud:2018sfi, CERN-EP-2018-139, Belanger:2013kya}. The high-luminosity run of the LHC (HL-LHC) has the potential to constrain all the Higgs couplings to an even greater precision~\cite{Peskin:2012we, Peskin:2013xra}. Besides, it also promises to shed light on the cubic and quartic couplings of the Higgs boson through its pair production.

On the other hand, the existence of dark matter in the universe has been repeatedly verified by astrophysical and cosmological observations ranging from galactic to cosmological scales. Apart from the fact that dark matter interacts gravitationally, the only quantitative aspect that we know about it is its relic abundance~\cite{Ade:2015xua}, $\Omega_{c}h^2=0.1199 \pm 0.0027$. However, its nature is still unknown and we expect it to be some electrically neutral particle with no colour quantum number. Amidst the various propositions, the Weakly Interacting Massive Particle (WIMP) stands out as one of the most attractive candidates by attributing to its simplicity and predictability. The observed relic abundance can be explained by the thermal freeze-out mechanism of the WIMP, when its mass is around the electroweak scale. An extension of the SM with a WIMP can help us understand better the origin of the electroweak symmetry breaking (EWSB). The predicted interactions of the WIMP with the SM particles greatly motivate the experimental community to search for this elusive particle at collider experiments, direct dark matter detection experiments at underground laboratories and from indirect detections from cosmological and astrophysical observations.

The lightest supersymmetric particle (LSP) has been considered as the most attractive candidate for cold dark matter due to the fact that the supersymmetric (SUSY) theories alleviate most of the aforementioned limitations faced by the SM. Unfortunately however, SUSY models are gradually getting severely constrained because of the lack of any evidence for superpartners. Lack of any conclusive signatures of WIMPs have gradually pushed the celebrated WIMP scenarios to the corner. In the present study we take recourse to one of the simplest models, the inert Higgs doublet model (IDM)~\cite{Deshpande:1977rw, Carmona:2015haa} which has an inbuilt WIMP candidate. The IDM is possibly the simplest limit of a general two Higgs doublet model, where the additional doublet consisting of complex scalar fields only couples to the SM Higgs and gauge bosons and not to the fermions. However, the most interesting aspect of this model is that the additional doublet is odd under a $\mathbb{Z}_2$ symmetry rendering the occurrence of an even number of \textit{inert} particles at any interaction vertex. It has also been shown that the neutral scalar or pseudoscalar in the additional doublet can be considered as WIMPs and hence as a viable cold dark matter candidate in the universe~\cite{Ma:2006km,Barbieri:2006dq}. In this model, obtaining the correct relic abundance does not require a fine tuning but only requires adjusting its couplings or through co-annihilation with another particle~\cite{LopezHonorez:2006gr}. Several experiments like LUX~\cite{Akerib:2013tjd}, SuperCDMS~\cite{Agnese:2014aze}, Fermi-LAT~\cite{Ackermann:2010ij, FermiLAT:2011ab}, AMS-02~\cite{Aguilar:2014mma,Aguilar:2013qda} and very recently XENON 1T~\cite{Aprile:2012zx, Aprile:2017iyp, Aprile:2018dbl}, have tested the dark matter scenario in the context of WIMP searches. These direct dark matter searches have now constrained the mass of the dark matter candidate in the IDM to around half of the mass of the SM Higgs boson (125 GeV) or above $\sim 500$ GeV~\cite{Goudelis:2013uca,Blinov:2015qva,Diaz:2015pyv}. The resonance mass around $\sim 62$ GeV might still not be completely excluded in the future~\cite{Abe:2015rja} by direct detection experiments like LZ~\cite{Akerib:2015cja}. Outside these ranges, the dark matter candidate can only contribute to a fraction of the total thermal relic density. Because of its simplicity and richness, the IDM has been exhaustively studied in astrophysical and cosmological studies~\cite{Gustafsson:2007pc, Agrawal:2008xz, Andreas:2009hj, Nezri:2009jd, Arina:2009um, Gong:2012ri} and studies pertaining to collider physics~\cite{Gustafsson:2012aj, Lundstrom:2008ai, Cao:2007rm, Dolle:2009ft, Miao:2010rg, Wang:2012zv, Osland:2013sla, Belyaev:2016lok}. There have been several studies in the context of the LHC which considers the Drell-Yan production of $HA$ or $H^+H^-$~\cite{Arhrib:2013ela, Ilnicka:2015ova, Poulose:2016lvz} and then further decays of $H \to A Z^{(*)},H^{\pm} \to A W^{\pm(*)}$ to yield a final state of dileptons or dijets + $\slashed{E}_T$. Here, $H$ and $A$ are respectively the additional neutral scalar and pseudoscalar and $H^{\pm}$ is the charged scalar, in this doublet. Depending on the $H-H-h$ coupling, a monojet signature can also be looked for at the LHC by a pair production of these scalars, if $H$ is the dark matter candidate. To give a broad picture, IDM connects the Higgs to dark matter by acting as a portal between the visible and the invisible sector. The on-shell corrections to $hVV, hff$ and $hhh$ couplings have been obtained in Refs.~\cite{Arhrib:2015hoa,Kanemura:2016sos} after considering constraints from perturbative unitarity, vacuum stability and relic abundance. The corrections have been shown to be substantial and can be $\simeq$100\% in the regime of light dark matter masses. These calculations could be of immense importance when the experiments start to constrain the cubic and quartic higgs self-couplings more precisely. Moreover, it has been shown in Ref.~\cite{Klasen:2013btp} that the electroweak corrections to the direct detection cross-sections in the IDM can be substantial.

In the present work, we revisit the electroweak correction of the $H-H-h$ vertex and show its effects on the relic abundance calculation and the direct detection cross-section. We are guided by the principle that in order to ascertain the importance of any model for dark matter searches, one needs to look at the higher order corrections which might lead to a significant shift in the parameter space under the constraints from relic density, direct detection cross-section, oblique corrections and collider limits. IDM is one of the simplest models to test our claim and through this we show the importance of precision measurements in the dark matter sector.
  
We organise the paper as follows. In section~\ref{sec2}, we briefly sketch the inert doublet model and its important aspects. We outline the constraints coming from perturbativity, vacuum stability etc. in section~\ref{sec3}. We then discuss our renormalisation procedure in section~\ref{sec4} but leave all the details in Appendices~\ref{appA} and~\ref{appB}. In section~\ref{sec5}, we discuss the numerical results and finally we summarise and conclude in section~\ref{sec6}.

\section{A Brief Review of the inert doublet model}
\label{sec2}

In addition to the SM fields, the inert doublet model employs an additional scalar doublet, $\Phi_2$. Moreover, the framework is endowed with a global $\mathbb{Z}_2$ symmetry under which $\Phi_2$ has a negative charge whereas the SM fields have a positive charge. The most general renormalisable scalar potential involving two doublets is then given by~\cite{Kanemura:2016sos}
\begin{eqnarray}
V &=& \mu_1^2{\Phi^\dagger_1}{\Phi_1} + \mu_2^2{\Phi^\dagger_2}{\Phi_2} +
\frac{\lambda_1}{2}({\Phi^\dagger_1}{\Phi_1})^2+\frac{\lambda_2}{2}({\Phi^\dagger_2}{\Phi_2})^2 \nonumber \\
&+&
\lambda_3({\Phi^\dagger_1}{\Phi_1})({\Phi^\dagger_2}{\Phi_2}) 
+\lambda_4({\Phi^\dagger_2}{\Phi_1})({\Phi^\dagger_1}{\Phi_2})
+ \Big[\frac{\lambda_5}{2}({\Phi^\dagger_1}{\Phi_2})^2 + {\rm h.c.}\Big],
\label{potential}
\end{eqnarray}
where all parameters are real, and $\Phi_1$ is the SM Higgs doublet. Noting that the $\mathbb{Z}_2$ symmetry prevents $\Phi_2$ from picking a vacuum expectation value (\textit{vev}), enables us to parametrise the doublets directly in terms of the physical scalars as
\begin{eqnarray}
\Phi_1 = \left(\begin{array}{c}
G^+ \\ \frac{1}{\sqrt{2}}(v + h + iG) \end{array}\right) ~{\rm ~and}~~ 
\Phi_2 = \left(\begin{array}{c}
H^+ \\ \frac{1}{\sqrt 2}(H + iA) \end{array}\right)   
\label{fields}
\end{eqnarray}

The charge of $\Phi_2$ under the $SU(2)_L \times U(1)_Y$ gauge group is (2,$\frac{1}{2}$) irrespective of the values chosen for $\l_4$ and $\l_5$. Therefore, it indeed has interactions with the gauge bosons of the $\Phi \Phi V V$ and $\Phi \Phi V$ forms, where, $\Phi = H, A, H^+$ and $V =W^{\pm}, Z$. A $\Phi V V$ vertex, however, is disallowed by the $\mathbb{Z}_2$ symmetry. The masses are calculated to be
\bea
m^2_h = \l_1 v^2, \; m^2_{H^{\pm}} = \mu_2^2 + \frac{1}{2}\lambda_3v^2, \; m^2_{H} = \mu_2^2 + \frac{1}{2}\lambda_L v^2, 
\; m^2_{A} = \mu_2^2 + \frac{1}{2}\lambda_A v^2, 
\label{eq:scamass}
\eea
where $\lambda_{L/A} = (\lambda_3 + \lambda_4 \pm \lambda_5)$. Besides, $\lambda_1 = \frac{m^2_h}{v^2}$ is determined using $m_h = 125$ GeV.\\

We choose $h$ to be the SM-like Higgs with mass $\sim$ 125 GeV. It is easily seen that $H,A,H^{\pm}$ are rendered stable by the $\mathbb{Z}_2$ symmetry and thus consequently, $H$ and $A$ are potential candidates for DM. While a detailed account of DM phenomenology for the IDM can be found in~\cite{Gustafsson:2012aj, Lundstrom:2008ai, Cao:2007rm, Dolle:2009ft, Miao:2010rg, Wang:2012zv, Osland:2013sla,  Belyaev:2016lok}, a few statements are still in order. Relic abundance in the PLANCK ballpark is achieved in the two mass regions (a) $50 \lesssim m_{DM} \lesssim 80$ and (b) $m_{DM} \gtrsim 500$ GeV. In region (a), annihilation dominantly proceeds through the exchange of an $s$-channel $h$. The sub-dominant contribution to the relic density comes from the $t$-channel processes to vector boson final states mediated by $A$ and $H^\pm$. On the other hand, one must have $m_H \simeq m_A \simeq m_{H^{\pm}}$ in order to generate $\Omega_{\rm DM}h^2$ $\simeq$ 0.1 in region (b). Co-annihilation thus becomes inevitable in this case.

In this study, $H$ is chosen to be the DM candidate. A crucial observation that emerges is, in region (a), $\Omega_{\rm DM}h^2$ is highly sensitive to the value of the $H-H-h$ trilinear coupling which is $-\l_Lv$ at LO with $\l_L = \l_3 + \l_4 + \l_5$. One therefore  expects the region (a) to be naturally more sensitive to the aforementioned radiative effect. This motivates one to review the entire phenomenology by incorporating radiative corrections to, if not all parameters, to the $H-H-h$ portal interaction nonetheless. In addition, beyond the leading order, the parameters that do not participate in the tree level phenomenology of region (a) (such as $\l_2$ and masses of the CP-odd and charged scalars), will now have their respective roles in the ensuing quantum effects. 
 
\section{Constraints}
\label{sec3}

\begin{figure} 
\begin{center}
\includegraphics[scale=0.30]{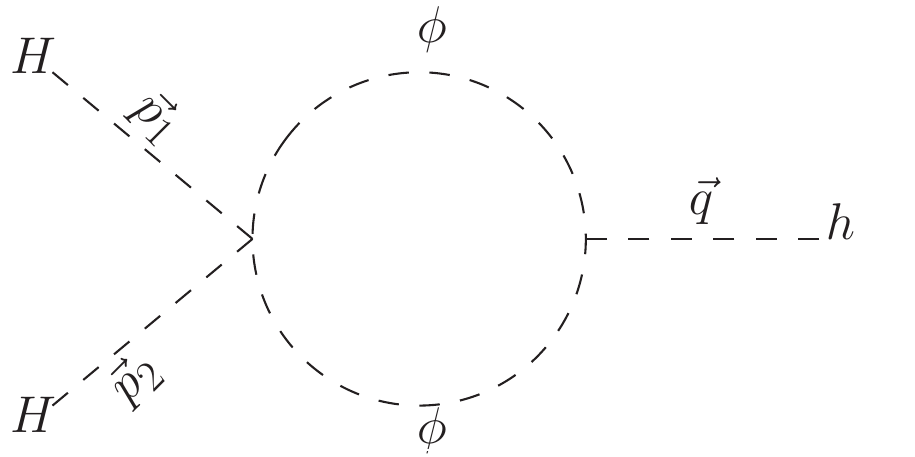}~~~~~~~\includegraphics[scale=0.30]{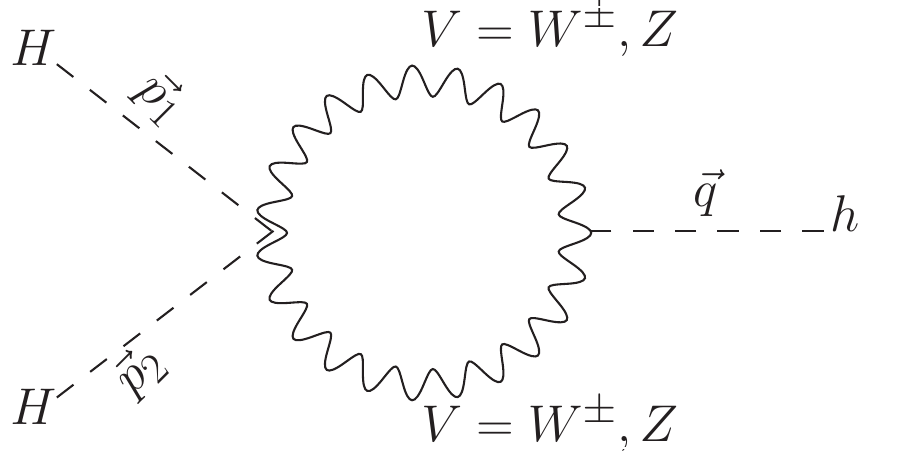}\\
\smallskip
\includegraphics[scale=0.30]{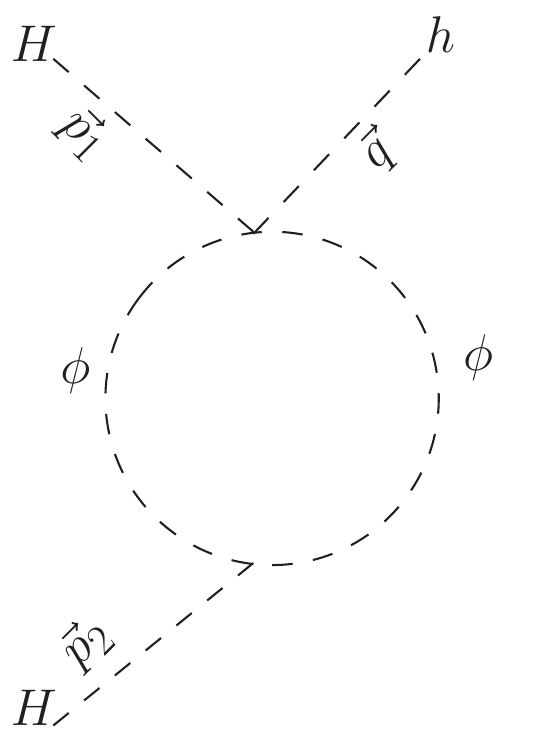}~~~~~~~\includegraphics[scale=0.30]{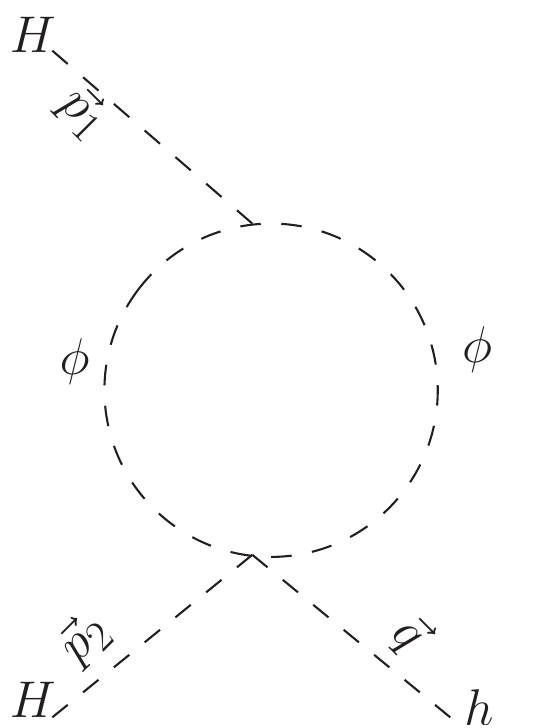}\\
\smallskip
\includegraphics[scale=0.30]{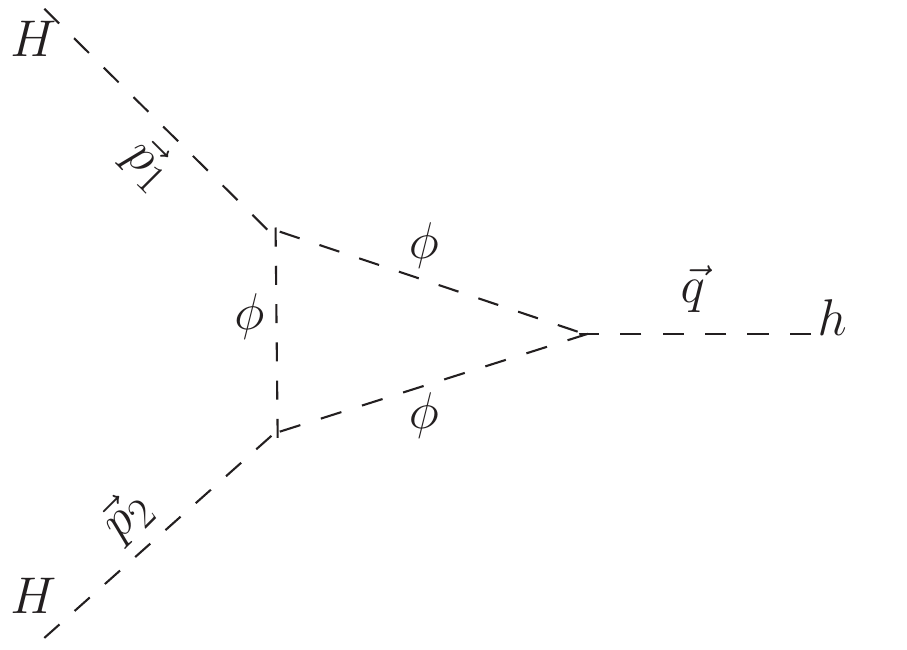}~~~~~~~\includegraphics[scale=0.30]{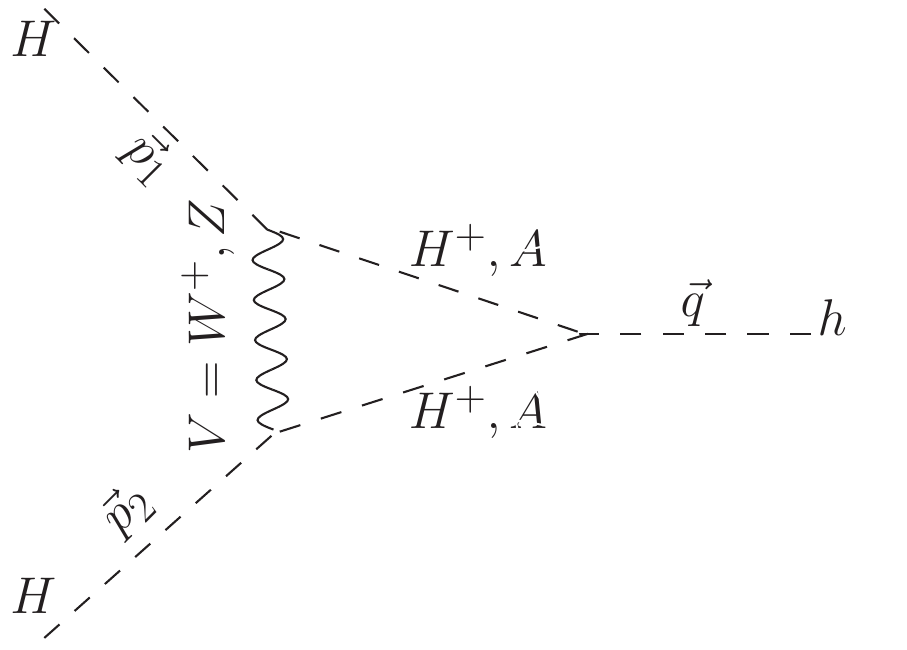}~~~~~~~\includegraphics[scale=0.30]{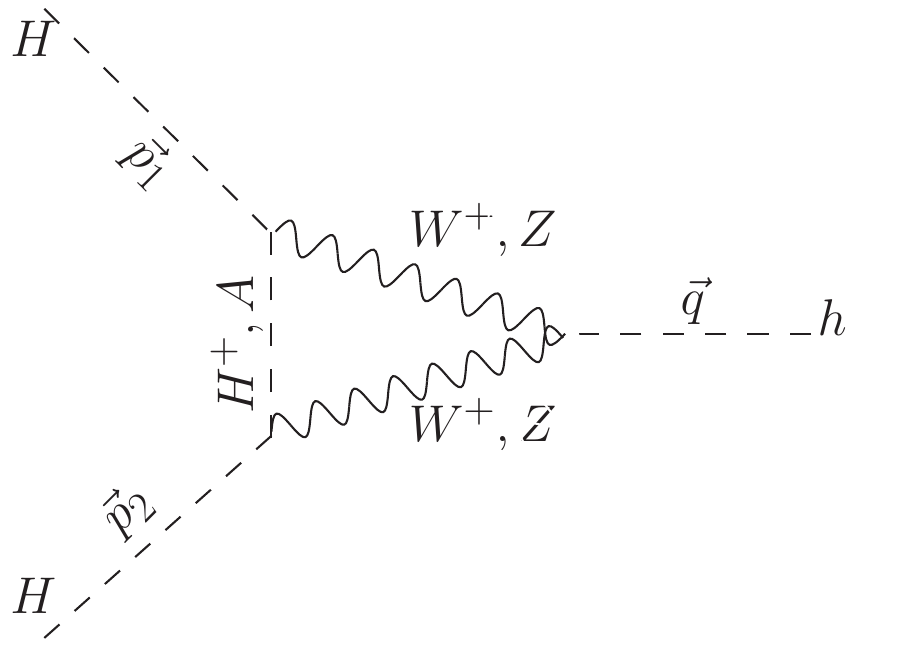}
\caption{Figure showing representative one-loop Feynman diagrams~\cite{Binosi:2008ig} for the $H-H-h$ trilinear vertex. Here, $\phi$ is used to denote $h, H, A, H^+, G^0, G^+$ or a subset that preserves the $\mathbb{Z}_2$ symmetry (even number of particles from $\Phi_2$) in each vertex.}
\label{fig:Feynman}
\end{center}
\end{figure}

Our goal is to take a recourse to the DM phenomenology in the IDM after carrying out one-loop corrections to the $H-H-h$ coupling (the Feynman diagrams for the three point function are shown in Fig~\ref{fig:Feynman}). In the process, we obey various constraints stemming from both theory and experiments. On the theoretical side, perturbativity, unitarity and vacuum stability can appreciably constrain an extended Higgs sector as in the IDM. From perturbativity, we impose the constraints $|\l_i| \leq 4\pi$, for $i = 1,2,..5$. The $2\rightarrow 2$ matrix element corresponding to the scattering of the longitudinal components of the gauge bosons can be mapped to a corresponding matrix for the scattering of the goldstone bosons~\cite{Lee:1977eg, Akeroyd:2000wc, Horejsi:2005da, Gorczyca:2011he}. The theory respects unitarity if the absolute value of each eigenvalue of the aforementioned amplitude matrix does not exceed 8$\pi$.

The tree level potential remains positive definite along various directions in the field space if the following conditions are met~\cite{Kanemura:2016sos},

\bea
\label{e:vsc1}
\lambda_{1} > 0, \nonumber \\
\label{e:vsc2}
\lambda_{2} > 0, \nonumber \\
\label{e:vsc3}
\lambda_{3} + \sqrt{\lambda_{1} \lambda_{2}} > 0, \nonumber \\
\label{e:vsc4}
\lambda_{3} + \lambda_{4} - |\lambda_{5}| + \sqrt{\lambda_{1} \lambda_{2}} > 0 
\eea
\label{eq:vsc}

On the experimental side, the bounds on the oblique parameters are taken into account. The most prominent of these is the constraint on the $T$ parameter which puts restriction on the mass splitting between the $\mathbb{Z}_2$ odd scalars. The contribution coming from the inert scalars can be expressed as~\cite{Grimus:2008nb} follows.

\bea
\Delta T &=& \frac{g^2}{64 \pi^2 m^2_W \a}
\Big(F(m^2_{H^+},m^2_H) + F(m^2_{H^+},m^2_A) - F(m^2_H,m^2_A)\Big) \nonumber \\
\text{where}~F(x,y) &=& \frac{1}{2}(x + y) - \frac{x y}{x - y}\text{log}\Big(\frac{x}{y}\Big)
\eea

We use the NNLO global electroweak fit results obtained by the Gfitter group~\cite{Baak:2014ora},
\bea
\Delta S = 0.05 \pm 0.11, \; \Delta T = 0.09 \pm 0.13 
\eea

In the absence of any mixing between $h$ and the $\mathbb{Z}_2$ odd scalars, the tree level couplings of $h$ with the fermions and gauge bosons remain unaltered with respect to their SM values. This implies that the production cross sections of $h$ at the LHC in case of the IDM do not change \emph{w.r.t.} the corresponding SM values. And in case of an unaltered $h$-production cross section, the signal strength in the diphoton channel becomes $R_{\g \g} = \frac{\Gamma_{h \to \g \g}^{\text{IDM}}}{\Gamma_{h \to \g \g}^{\text{SM}}}$. The charged Higgs $H^+$ coming from the inert doublet leads to an additional one-loop term in the $h \to \g \g $ amplitude~\cite{Swiezewska:2012eh}. That is,

\bea
\mathcal{M}^{\text{IDM}}_{h \to \g \g} &=& 
\frac{4}{3}A_f\Big(\frac{m^2_h}{4 m^2_t}\Big)
 + A_V\Big(\frac{m^2_h}{4 m^2_W}\Big)
 + \frac{\l_3 v^2}{2 m^2_{H^+}} A_S\Big(\frac{m^2_h}{4 m^2_{H^+}} \Big) \nonumber \\
\Gamma^{\text{IDM}}_{h \to \g \g} &=& \frac{G_F \a^2 m_h^3}{128 \sqrt{2} \pi^3} |\mathcal{M}^{\text{IDM}}_{h \to \g \g}|^2,
\eea
where $G_F$ and $\a$ denote respectively the Fermi constant and the QED fine-structure constant. The loop functions are listed below.
\bea
A_f(x) &=& \frac{2}{x^2}\big(x + (x -1)f(x)\big), \nonumber \\
A_V(x) &=& -\frac{1}{x^2}\big(2 x^2 + 3 x + 3(2 x -1)f(x)\big),\nonumber \\
A_S(x) &=& -\frac{1}{x^2}\big(x - f(x)\big), \nonumber \\
\text{with} ~~f(x) &=& \big(\text{sin}^{-1}\sqrt{x}\big)^2,
\eea

where $A_f(x), A_V(x)$ and $A_S(x)$ are the respective amplitudes for the spin-$\frac{1}{2}$, spin-1 and spin-0 particles in the loop and $x = m_h^2/4 m_{f/V/S}^2$. Therefore, ensuring $\mu_{\g\g}$ to lie within the experimental uncertainties, implies that the analysis respects the latest signal strength at 13 TeV from ATLAS~\cite{Aaboud:2018xdt} and CMS~\cite{CMS-PAS-HIG-18-029, Khachatryan:2016vau, Sirunyan:2018ouh}, \textit{viz.},
\bea
 \mu_{\gamma \gamma} &=& 0.99 \pm 0.14 \; \to \textrm{ATLAS combined} \nonumber \\
                     &=& 1.15 \pm 0.15 \; \to \textrm{CMS gluon fusion channel} \nonumber \\
                     &=& 0.8^{+0.4}_{-0.3} \; \to \textrm{CMS VBF channel}
\eea
Upon using the standard combination of signal strengths and uncertainties~\footnote{$$\frac{1}{\bar{\sigma}^2}=\sum_i \frac{1}{\sigma_i^2}, \; \textrm{and} \; \frac{\bar{\mu}}{\bar{\sigma}^2}=\sum_i \frac{\mu_i}{\sigma_i^2},$$ where $\mu_i$ and $\sigma_i$ are the individual signal strengths and their 1-$\sigma$ uncertainties respectively.}, we obtain $\mu_{\g\g} = 1.04 \pm 0.1$.
Furthermore, we also require the invisible branching ratio of the SM-like Higgs to be BR($h \rightarrow \rm inv$) $<$ 0.15. Lastly, the LEP exclusion limits have been imposed as $m_A \gsim 100$ GeV~\cite{Schael:2006cr} and $m_{H^+} \gsim 90$ GeV~\cite{Achard:2003gt, Lundstrom:2008ai}.

The IDM has been one of the most popular models that has garnered attention amidst astrophysicists and particle physicists alike, owing to its simplicity and predictive power. Apart from the modification of the diphoton partial width with respect to the SM expectation, the constraint on the $T$-parameter and the bound from the invisible decay of the SM-like Higgs boson, there are a multitude of search channels which have be used to constrain the IDM parameter space. In Refs.~\cite{Schael:2006cr, Achard:2003gt, Lundstrom:2008ai} it has been shown that the points satisfying the intersection of the following conditions
\be
m_H < 80 \; \textrm{GeV}, m_A < 100 \; \textrm{GeV} \; \textrm{and} \; m_A - m_H > 8 \; \textrm{GeV},
\ee
are excluded by the LEP II data as they would lead to a di-lepton/di-jet signature along with missing energy. Reference~\cite{Lundstrom:2008ai} showed this in the context of a reinterpretation of the second neutralino. This study has been studied for low values of $m_H$ and $m_A$ with a significant mass gap in the context of the LHC Run I data~\cite{Belanger:2015kga}. Moreover, Ref.~\cite{Pierce:2007ut} studied the process $e^+e^- \to H^+ H^-$ in the context of LEP II data and obtained a bound of $m_{H^{\pm}} > 70$ GeV. In Refs.~\cite{Ilnicka:2015jba, Belyaev:2016lok}, the authors study multifarious channels at the LHC in order to impose constraints on the IDM parameter space. Both these studies first look into the constraints ensuing from the SM-like Higgs mass, present limit on the SM-like Higgs width and the Higgs signal strengths. Recast of searches in mono-jet ($HH +$ jet and $HA +$ jet, $gg, gq$ and $q\bar{q}$ initiated), mono-$Z$ ($q\bar{q}$ initiated), mono-Higgs ($gg$ initiated with $HHh$ and $q\bar{q}$ initiated with $HAh$) and vector boson fusion ($HH +$ jets) were performed in Refs.~\cite{Ilnicka:2015jba, Belyaev:2016lok}. The mono-jet processes having the highest cross-section amongst the rest were considered for the 8 TeV and the projected 13 TeV scenarios in Ref.~\cite{Belyaev:2016lok}. The projected LHC 13 TeV study in Ref.~\cite{Belyaev:2016lok} allows small values of $\lambda_L$ between $m_H \in [50,80]$ GeV. Besides, from the same study, $m_H \in [55, 80]$ GeV is allowed when $m_A > 100$ GeV from the mono-jet requirements. Similar bounds are presented in the $m_A-m_{H^{\pm}}$ plane. As has been pointed out in Ref.~\cite{Eiteneuer:2017hoh}, the Galactic Centre Excess~\cite{TheFermi-LAT:2017vmf, TheFermi-LAT:2015kwa, Ackermann:2015lka} (GCE) best-fit data favours small values of $\lambda_L$ which is the driving coupling for the mono-$X$ like searches. Thus, LHC has a very low impact on constraining the parameter space of the IDM from such searches. However, there are other searches which are independent of the size of $\lambda_L$ and depend on the masses of the particles and the splitting. In Ref.~\cite{Belanger:2015kga}, the chargino and neutralino pair production processes have been studied in details. In particular, they focussed on $q\bar{q} \to AH \to Z^{(*)}HH \to \ell^+\ell^- HH, q\bar{q} \to Z \to H^{\pm}H^{\pm} \to W^{\pm(*)}W^{\mp(*)}HH \to HH \nu \bar{\nu} \ell^+ \ell^-, q\bar{q} \to Z H H \to \ell^+ \ell^- HH$ and $q\bar{q} \to Z \to Z h^{(*)} \to \ell^+ \ell^- HH$ where the first three processes are free from the $\lambda_L$ coupling and involve only the gauge couplings and are thus dependent only on the masses of the scalars. Even though there are no dedicated searches for the IDM from LHC, Ref.~\cite{Belanger:2015kga} recast a SUSY analysis involving di-leptons plus $\slashed{E}_T$. Their results show that $m_H$ up to 35 GeV are excluded at 95\% CL with $m_A \sim 100$ GeV. These limits were shown to get stronger with larger values of $m_A$ and the limits went up to $m_H \sim 45 \; (55)$ GeV for $m_A \sim 140 \; (145)$ GeV and $m_{H^{\pm}} \sim 85 \; (150)$ GeV. For a more detailed discussion, we refer the reader to the aforementioned references.

\section{Outline of renormalisation}
\label{sec4}

In this section, we present an outline of the renormalisation procedure adopted in~\cite{Kanemura:2016sos}. The absence of mixing between $h$ and the inert scalars simplifies the machinery to some extent compared to a general two-Higgs doublet model. The IDM scalar sector can be conveniently described using $\{m_h,v,\mu_2,m_H,m_A,m_{H^+},\l_2,T_h\}$ as independent parameters. Here $T_h$ denotes the tadpole parameter for $h$. The necessary counterterms are generated by shifting these parameters about their renormalised values as follows. 
\bea
m^2_h \rightarrow m^2_h + \delta m^2_h \nonumber \\
v \rightarrow v + \delta v \nonumber \\
\mu_2^2 \rightarrow \mu_2^2 + \delta \mu_2^2 \nonumber \\
m_H^2 \rightarrow m_H^2 + \delta m_H^2 \nonumber \\
m_A^2 \rightarrow m_A^2 + \delta m_A^2 \nonumber \\
m_{H^+}^2 \rightarrow m_{H^+}^2 + \delta m_{H^+}^2 \nonumber \\
\l_2 \rightarrow \l_2 + \delta \l_2
\eea

It is to be noted that we do not need to compute the shift in $T_h$ in our case. Moreover, the wave-function renormalisation is invoked as
\bea
\phi \rightarrow (1 + \frac{1}{2} \delta Z_{\phi}) \phi,
\eea
where $\phi = h, H , A , H^+$. A more detailed treatment of the renormalisation scheme can be found in~\cite{Kanemura:2016sos, PhysRevD.70.115002}. In the on-shell (OS) scheme, the mass and field shifts can be expressed in terms of the 1PI amplitudes as follows:

\bea
\delta m^2_{\phi} & = & \Pi_{\phi \phi} (m^2_{\phi}) \nonumber \\
\delta Z_{\phi} & = &  -\frac{d}{d p^2} \Pi_{\phi \phi} (p^2)|_{p^2 = m^2_{\phi}}.
\eea
\label{eq:delmh2zh}

These 1PIs are detailed in~\ref{appA} and~\ref{appB}. The quantity of central importance in this study is the renormalised $H-H-h$ form factor which replaces its tree level counterpart in dark matter calculations. We denote it by $\Gamma_{HHh}^{\rm ren}$ and decompose it as 
\bea
\Gamma_{HHh}^{\rm ren}(p^2_1,p^2_2,p^2) = \Gamma_{HHh}^{\rm tree} + \Gamma_{HHh}^{\rm 1PI}(p^2_1,p^2_2,p^2) + \delta \Gamma_{HHh}
\label{eq:ren}
\eea

Here, $p_1, p_2$ and $p = p_1 + p_2$ refer respectively to the momenta of the two annihilating $H$ and the $h$. On the right hand side of the equation, the first term refers to the tree level form factor. The second and the third terms respectively denote the unrenormalised 1PI amplitude at the one-loop level; and the corresponding counterterm. It is necessary to express the tree level form factor in terms of the independent parameters in order to generate the corresponding counterterm.

\bea
\Gamma_{HHh}^{\rm tree} &=& -\frac{2}{v}(m^2_H - \mu_2^2) \\
\rm Leading~ to  \nonumber \\
\delta \Gamma_{HHh} &=& -\frac{2(m^2_H - \mu_2^2)}{v}\Big[\frac{\delta m^2_H - \delta \mu_2^2}{m^2_H - \mu_2^2} - \frac{\delta v}{v} + \frac{1}{2} \delta Z_h + \delta Z_H\Big]
\label{eq:delmu2}
\eea

Now, we have fixed the counterterms $\delta m^2_H$, $\delta Z_h$ and $\delta Z_H$ from Eq.~\ref{eq:delmh2zh}. We also know the expression of $\delta v$ from SM. The only ambiguity in order to fix $\delta \Gamma_{HHh}$ is $\delta \mu^2_2$. There can be several ways to fix this counterterm~\footnote{In an ongoing work with F. Boudjema, G. Chalons and S. Hao, we are working on the full renormalisation of the IDM using several renormalisation schemes.}. What we choose in the present paper is demand that a physical quantity, here the decay width $h \rightarrow H H$, does not deviate \textit{w.r.t} its tree level value upon including one-loop corrections. This implies~\footnote{ The on-shell $h \to H H$ does not open up whenever $m^2_H > m^2_h/4$. So in Eqs.~\ref{eq:delmu2ct},~\ref{eq:delmu2ct2} and in the last term of 
\ref{eq:ren2}, we set $m^2_H = m^2_h/4$.} 

\begin{equation}
 \delta \Gamma_{HHh} = - \Gamma_{HHh}^{\rm 1PI}(m^2_H,m^2_H,m^2_h)
\label{eq:delmu2ct} 
\end{equation}

and 

\begin{equation}
 \delta \mu^2_2 = -\Bigg[\Bigg(\frac{\Gamma_{HHh}^{\rm 1PI}(m^2_H,m^2_H,m^2_h) v}{2(m^2_H - \mu^2_2)} + \frac{\delta v}{v} - \frac{1}{2} \delta Z_h - \delta Z_H\Bigg)(m^2_H - \mu^2_2) - \delta m^2_H\Bigg].
\label{eq:delmu2ct2}
\end{equation}

Upon considering all these counterterms, we get an effective correction of the form

\bea
\Gamma_{HHh}^{\rm ren}(p^2_1,p^2_2,p^2) = \Gamma_{HHh}^{\rm tree} + \Gamma_{HHh}^{\rm 1PI}(p^2_1,p^2_2,p^2) - \Gamma_{HHh}^{\rm 1PI}(m^2_H,m^2_H,m^2_h)
\label{eq:ren2}
\eea

The counterterms for the independent parameters are thus fixed. The quantity directly entering into our analysis is $\Gamma^{\text{ren}}_{H H h}(p_1^2,p_2^2,p^2)$. The expressions for the various two- and three-point 1PI amplitudes are relegated to appendices \ref{appA} and~\ref{appB}. 

In passing, we remark that an alternate way of fixing 
$\delta \mu_2^2$ is to assume that the SM gauge symmetry is unbroken, compute directly
the one-loop correction to $\mu_2^2$ itself and finally define $\delta \mu_2^2$ to be the UV-divergent part of
the same. This particular way of fixing this counterterm is therefore similar to what is done in the
 $\overline{\text{MS}}$
scheme. Hence, a dependence on the renormalisation scale
(say $\mu$) is expected.

\section{Numerical Results}
\label{sec5}

The quantity directly entering our analysis is $\Gamma^{\text{ren}}_{H H h}(p_1^2,p^2_2,p^2)$. Moreover, since dark matter particles annihilate manifestly in an on-shell fashion, we take $p_1^2 = p_2^2 = m_H^2$. Cold dark matter particles are non-relativistic, and this allows to write $p^2 = 4 m^2_H$~\footnote{We must note that even upon considering $p^2$ off-resonance will yield the same result because the vertex correction will be a function $f(p^2-m^2_h)$ which will partially cancel with the propagator, $\sim \frac{1}{p^2-m^2_h}$, and the correction will be a ``constant'' one.}~\footnote{The consequences of $p^2 \neq 4 m^2_H$ too are examined in the following section}. This allows us to treat the quantity $\l_L v$ + $\Gamma^{\text{ren}}_{H H h}( m_H^2, m_H^2, 4 m_H^2)$ as an ``effective" coupling in the present analysis. 

Coming to the analysis, we choose the following values chosen for the SM parameters, $m_h = 125.0$ GeV, $m_t = 173.2$ GeV, $m_b = 4.7$ GeV, $m_W = 80.3$ GeV and $m_Z = 91.2$ GeV~\footnote{Our mass choices lie in the 1-$\sigma$ uncertainty range of the measured Higgs and top masses from the LHC Run I and Run II data~\cite{Aad:2015zhl, Tanabashi:2018oca, Aaboud:2018wps, Sirunyan:2017exp}.}. The model points are sampled randomly through a scan of the parameter space within the ranges specified below.
\bea
\rm \mu^2_2 \in  [0 ~GeV^2,10^6 ~GeV^2] \nonumber \\
m_{A} \in  [100 ~\rm GeV, 500 ~\rm GeV]  \nonumber \\
m_{H^+} \in  [100 ~\rm GeV, 500 ~\rm GeV] \nonumber 
\eea

We avoid choosing $\mu^2_2 < 0$ in order to prevent the inert doublet from picking a \textit{vev}, alongside obeying the vacuum stability, perturbativity and unitarity constraints described in section~\ref{sec3}. Since the DM self-interaction $\l_2$ cannot be directly constrained, we choose $\l_2 = 0.1, 1$ and 5.0 in our scans. We also remind the readers that the upper bound on the invisible branching fraction of $h$ remains an important constraint whenever $m_H < m_h/2 $. 
The aforementioned constraint on the Higgs invisible branching ratio leads to $|\l_L v$ + $\Gamma^{\text{ren}}_{H H h}| < 0.05$ for $m_H = 55$ GeV for instance, and a tighter bound for a lower mass. We focus on the regions $50 ~\text{GeV} < m_H < 80 ~\text{GeV}$ and $m_H > ~500$ GeV (see section~\ref{sec2}) to illustrate our results.

\subsection{$50 ~\text{GeV} < m_H < 80 ~\text{GeV}$}


\begin{table}[h]
\centering
\begin{tabular}{||c c c c c c||}
\hline
\hline
Benchmark & $m_A$   &    ~~~~~~ $m_{H^+}$ &  ~~~~~~  $\l_L$ & ~~~~~~~~$R_{\g\g}$ 
& ~~~~~~~$\Delta T$ \\ \hline
BP1a      & 100 GeV &    ~~~~~~ 110 GeV   &  ~~~~~~  0.001 & $\in$ 
[0.904, 0.942] & $\in$ 
[0.005,0.010]\\ \hline
BP1b      & 200 GeV &    ~~~~~~ 210 GeV   &  ~~~~~~  0.001 & $\in$ 
[0.902, 0.911] & $\in$ 
[0.021,0.025]\\ \hline
BP2a      & 100 GeV &    ~~~~~~ 110 GeV   &  ~~~~~~ -0.001 & $\in$ [0.904, 0.943] 
& $\in$ 
[0.005,0.010]\\ \hline
BP2b      & 200 GeV &    ~~~~~~ 210 GeV   &  ~~~~~~ -0.001 & $\in$ [0.902, 0.911] 
& $\in$ 
[0.021,0.025]\\ \hline
BP3a      & 100 GeV &    ~~~~~~ 110 GeV   &  ~~~~~~  0.002 & $\in$ [0.904, 0.942] 
& $\in$ 
[0.005,0.010]\\ \hline
BP3b      & 200 GeV &    ~~~~~~ 210 GeV   &  ~~~~~~  0.002 & $\in$ [0.902, 0.911] 
& $\in$ 
[0.021,0.025]\\ \hline
BP4a      & 100 GeV &    ~~~~~~ 110 GeV   &  ~~~~~~ -0.002 & $\in$ [0.905, 0.943] 
& $\in$ 
[0.005,0.010]\\ \hline
BP4b      & 200 GeV &    ~~~~~~ 210 GeV   &  ~~~~~~ -0.002 & $\in$ [0.902, 0.911] 
& $\in$ 
[0.021,0.025]\\ \hline
\hline
\end{tabular}
\caption{Benchmark points, $50 ~\text{GeV} < m_H < 80 ~\text{GeV}$, satisfying the constraints listed in Section~\ref{sec3}, chosen to illustrate the effect of the one-loop corrections. The same benchmarks can also be expressed in terms of the corresponding $\mu_2$ values using $m^2_H = \mu_2^2 + \frac{1}{2}\l_L v^2$. 
The corresponding variations of $\Delta T$ and $R_{\g\g}$ are also indicated.}
\label{tab:BP}
\end{table}

\begin{figure} 
\begin{center}
\includegraphics[scale=0.38]{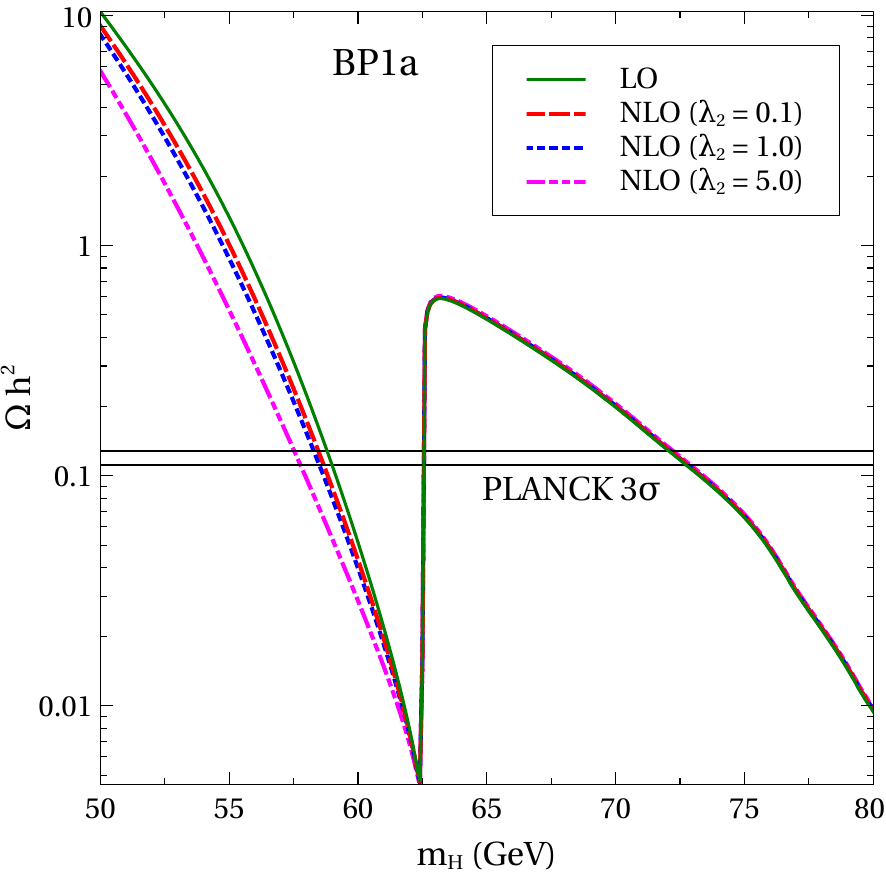}~~~ 
\includegraphics[scale=0.38]{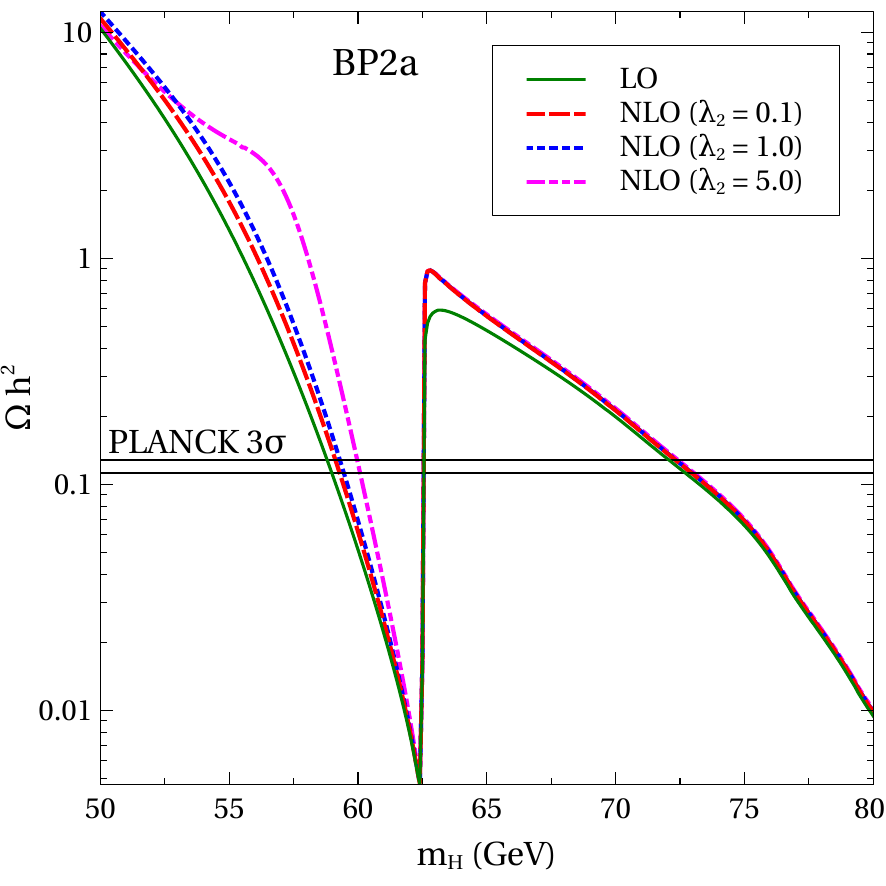}
\caption{Variation of the thermal relic with the DM mass (50 GeV $< m_H < $ 80 GeV). The tree level and one-loop values are denoted by the green solid and red dashed ($\l_2 = 0.1$), blue dotted ($\l_2 = 1$) and pink dot-dashed ($\l=5$) respectively.}
\label{full}
\end{center}
\end{figure}

We propose the following benchmarks in Tab.~\ref{tab:BP} that are combined with $\l_2 = 0.1, 1, 5$ discussed before. In addition to the constraints discussed in the previous section, such choices are also guided by a somewhat conservative requirement of $|\l_{3,4,5}| < 2$. Besides in Tab.~\ref{tab:BP}, we also show the constraints ensuing from the $T$-parameter and from $R_{\g\g}$. All the points abide by the $T$-parameter constraints mentioned in section~\ref{sec3}. As for $R_{\g\g}$, BP1a, BP1b, BP1c and BP1d are allowed within 1-$\sigma$ uncertainty of the combined signal strength mentioned above. For the remaining four benchmark points, the agreement is within 1.3-$\sigma$. In Fig.~\ref{full}, we show the full variation of $\Omega h^2$ as a function of $m_H$ for BP1a and BP2a. For all these benchmark points (listed in Tab.~\ref{tab:BP}), the relic density curve at the leading order cuts the $\Omega h^2 =$ 0.1 horizontal line at three distinct values of $m_H$. Out of these three values, the first two lie in the funnel region around $m_H \simeq m_h/2$ and the third around $m_H \sim 75$ GeV. The inert scalars participating in the loops can modify the tree level $H-H-h$ interaction strength considerably. Hence, significant quantitative deviations \emph{w.r.t.} the leading order calculations are noted. The primary features can be summarised as follows.

\begin{figure} 
\begin{center}
\includegraphics[scale=0.38]{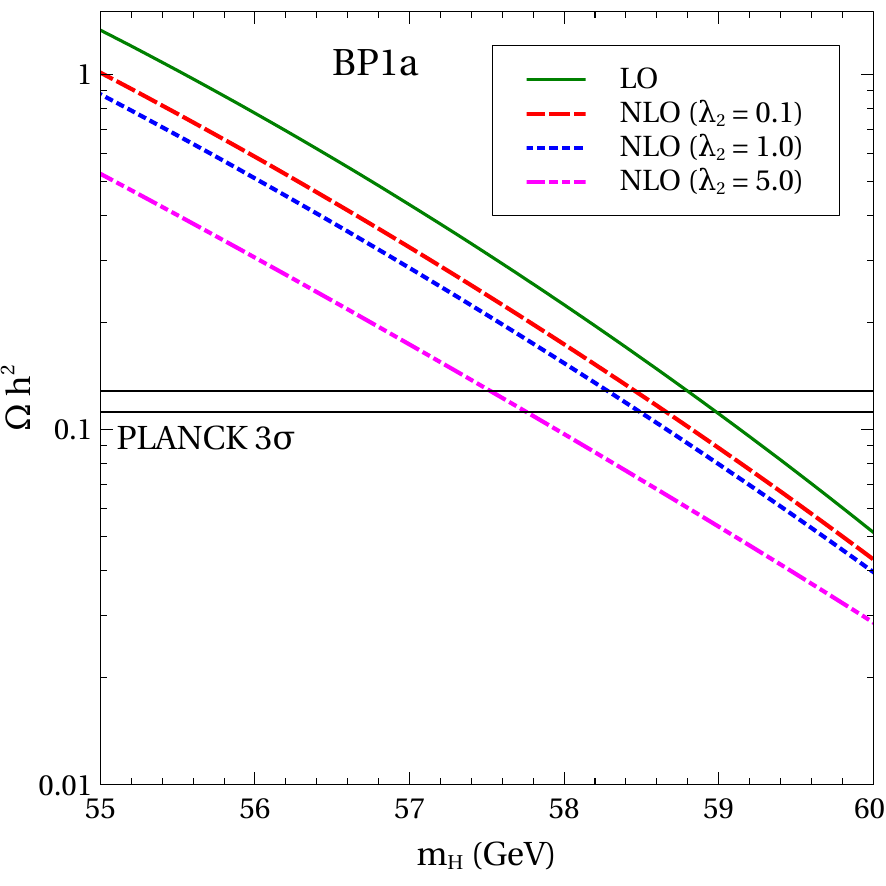}~~~ 
\includegraphics[scale=0.38]{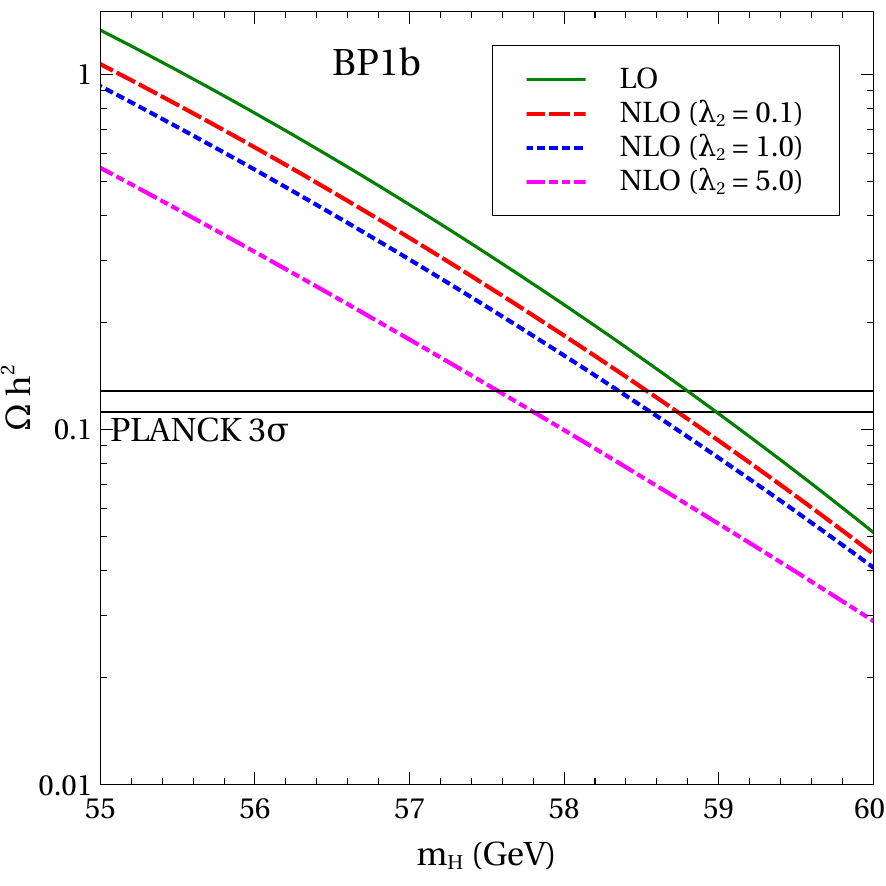}
\\
\includegraphics[scale=0.38]{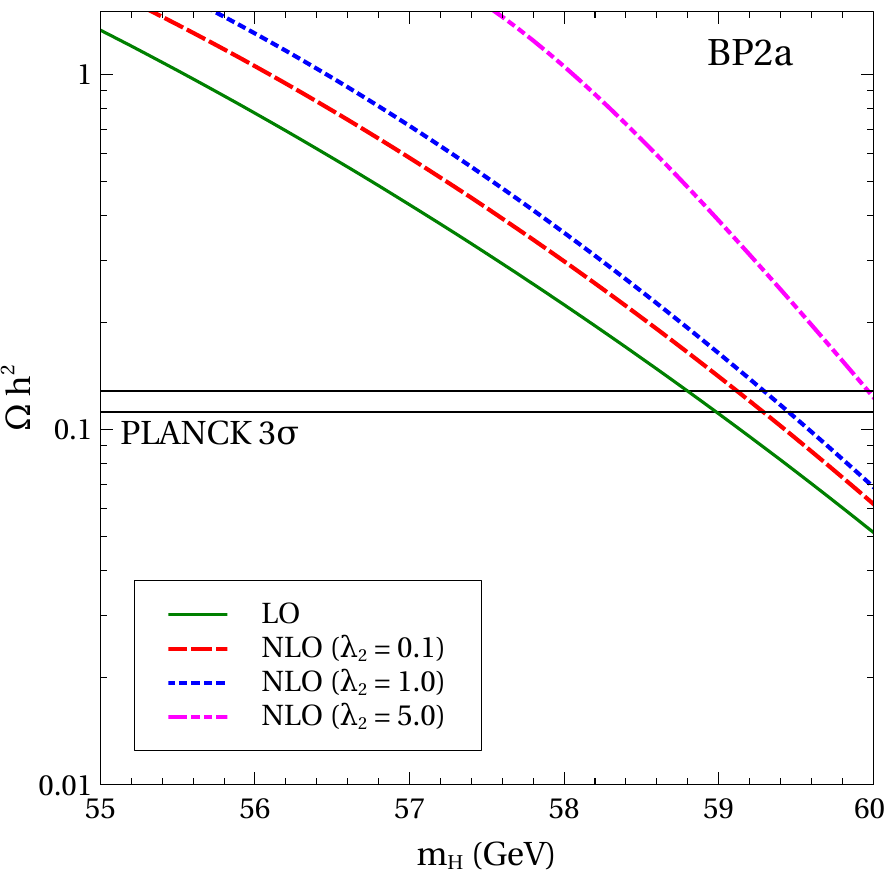}~~~ 
\includegraphics[scale=0.38]{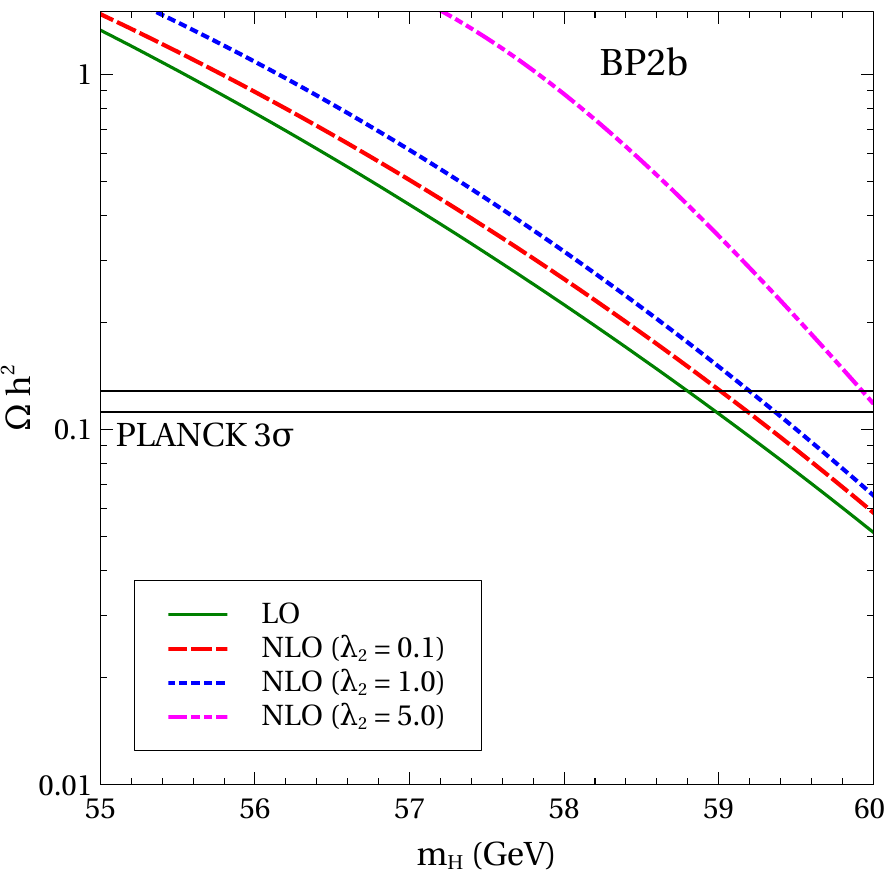}
\caption{Variation of the thermal relic with the DM mass (55 GeV $< m_H < m_h/2$). The tree level and one-loop values are denoted by the green solid and red dashed ($\l_2 = 0.1$), blue dotted ($\l_2 = 1$) and pink dot-dashed ($\l=5$) respectively.}
\label{relic1to2}
\end{center}
\end{figure}

Firstly, a flip in the sign of $\l_L$ does not imply a sign flip for the renormalised 1PI amplitude. This is because all the pre-factors of the loop functions (which also involve combinations of $\l_3, \l_4$ and $\l_5$ other than $\l_3 + \l_4 + \l_5$) do not reverse their signs at the same time. Secondly, $\Gamma^{\text{ren}}_{HHh}(m^2_H,m^2_H,4 m^2_H)$ is positive (negative) when $m_H$ is smaller (greater) than $m_h/2$. Such a crossover is expected since $\Gamma^{\text{ren}}_{HHh}(m^2_H,m^2_H,4 m^2_H)$ vanishes for $m_H = m_h/2$ (see Eq.~(\ref{eq:ren2})). This, in turn, stems from the fact that the particular choice of $\delta \mu^2$ considered here, that ultimately expresses $\Gamma^{\text{ren}}_{HHh}(m^2_H,m^2_H,4 m^2_H)$, is a difference of two 1PI form factors. Thirdly, the higher the value of $\l_2$, the higher is the magnitude of the deviation of the loop-corrected coupling \emph{w.r.t.} the corresponding tree level value. The aforementioned features are confirmed by an inspection of Fig.~\ref{relic1to2}, that zooms into the $m_H < m_h/2$ mass point around the funnel region, for BP1a, BP1b, BP2a and BP2b. In BP1a, a positive loop correction for $m_H < m_h/2$ adds to a positive $\l_L$ thereby further increasing the effective $H-H-h$ coupling in the same range of $m_H$ and ultimately lowering $\Omega h^2$. The highest value of $\l_2$ (= 5) thus corresponds to the curve with the lowest relic for this benchmark given that a larger $\l_2$ brings in a larger radiative correction.

\begin{figure} 
\begin{center}
\includegraphics[scale=0.38]{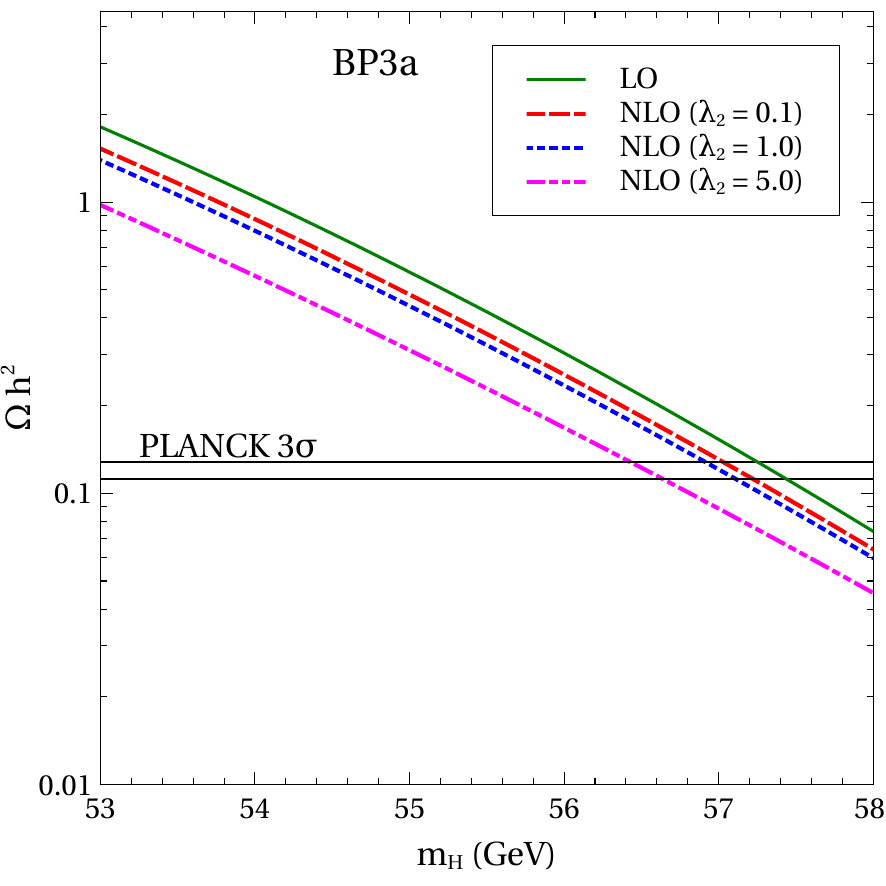}~~~ 
\includegraphics[scale=0.38]{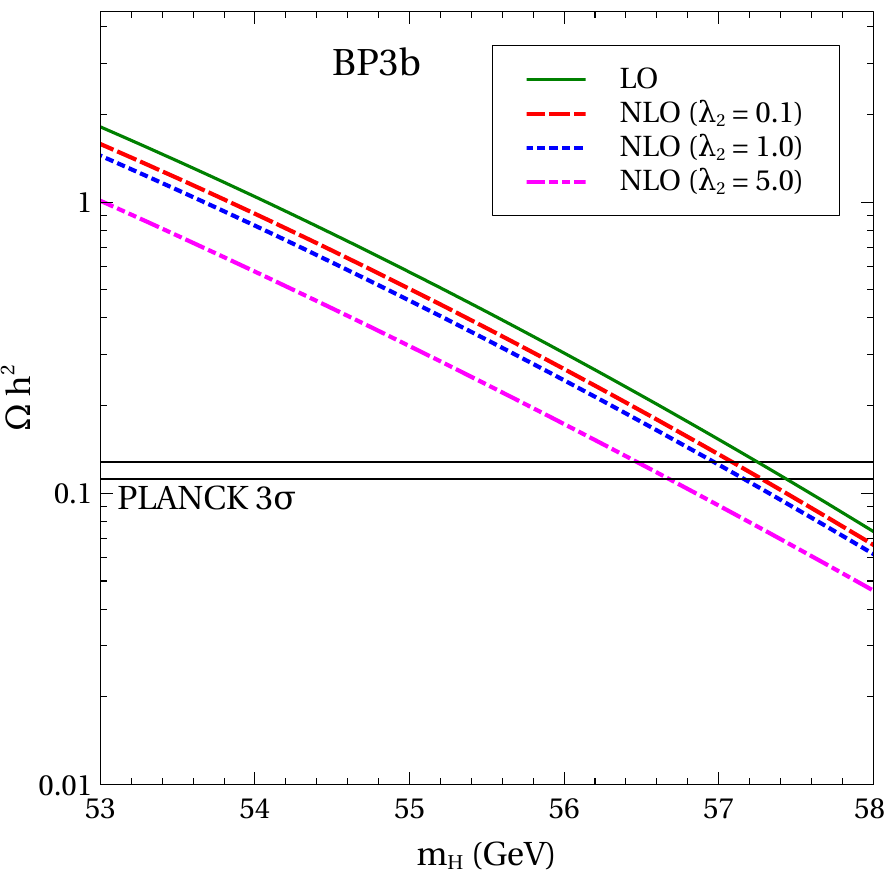}
\\
\includegraphics[scale=0.38]{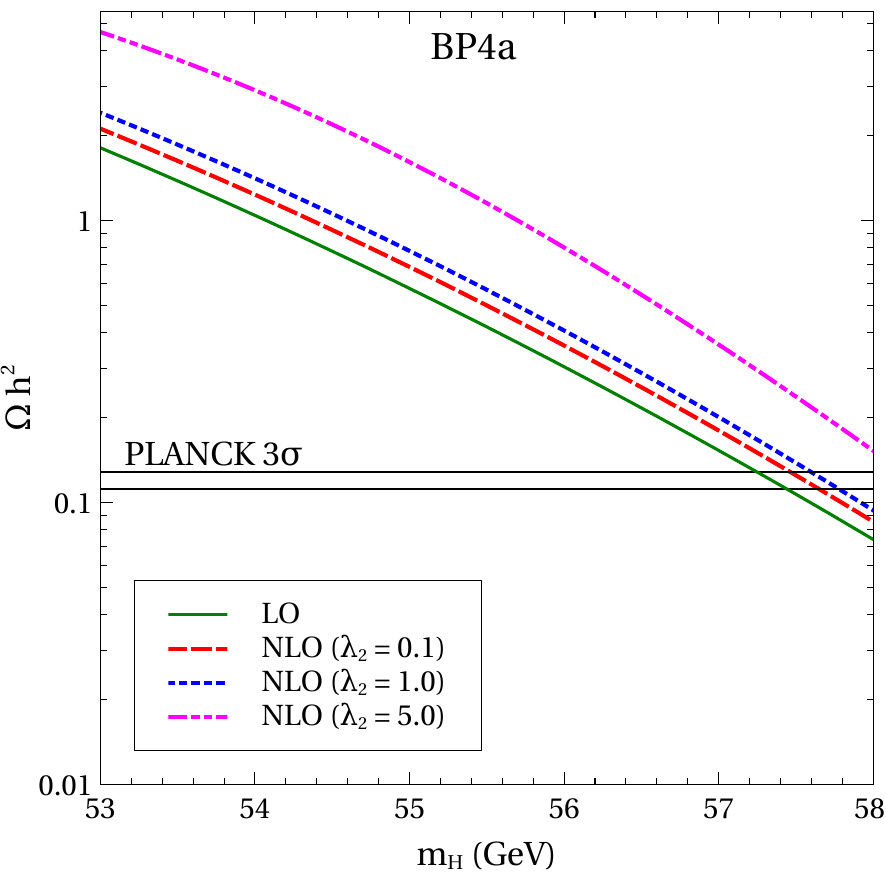}~~~ 
\includegraphics[scale=0.38]{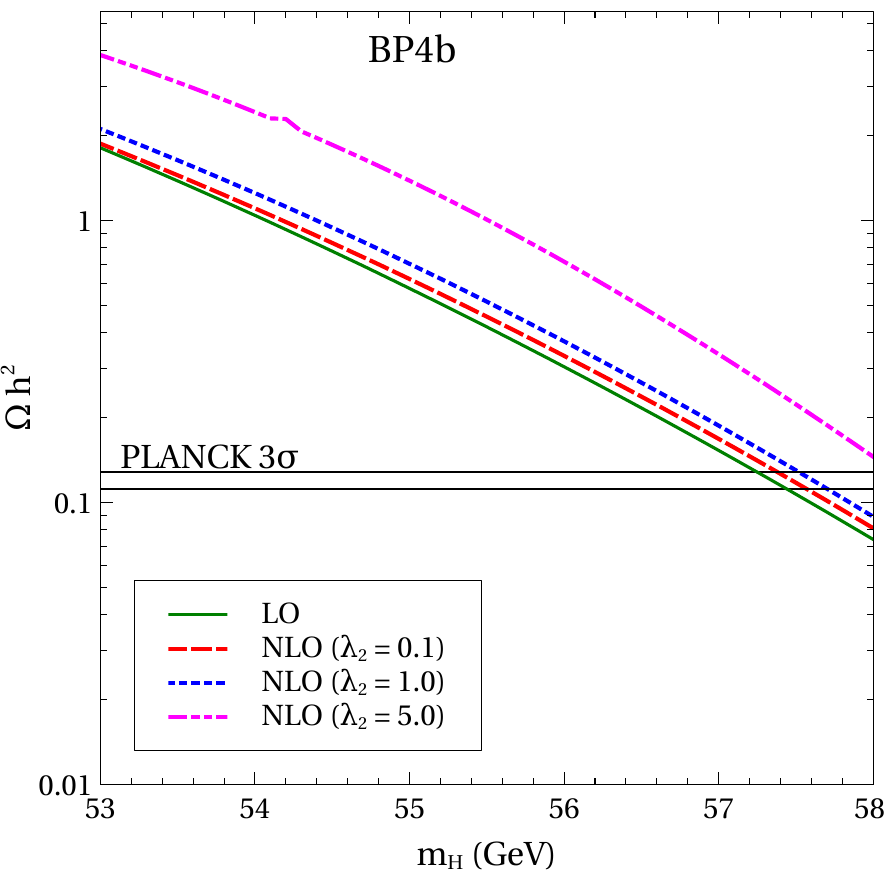}
\caption{The same as in Fig.~\ref{relic1to2} for BP3a, BP3b, BP4a and BP4b.}
\label{relic3to4}
\end{center}
\end{figure}

On the other hand, a positive loop correction adds to a negative $\l_L$ in case of BP2a, thereby reducing the effective coupling strength. The ordering of the tree level and loop-corrected relic curves in this case is therefore opposite to what is seen for BP1a. A suppression in loop correction with an increase in $m_A$ and $m_{H^+}$ with $m_H$ held fixed, is noted. One can confirm this upon inspecting Fig.~\ref{relic1to2}, where BP1b exhibits a smaller loop correction compared to BP1a. For instance, the shifts to the tree level interaction for $m_H = 55$ GeV in BP1a (BP1b) for $\l_2 = 0.1, 1$ and 5 respectively read $-24.06 \; (-15.65)\%, \; -33.99 \; (-26.86)\%$ and $-60.59 \; (-56.92)\%$. This is an important aspect of the \emph{non-decoupling} of the $m_H < 80$ GeV region, where all component scalars of the extra doublet are not simultaneously heavy. The various features discussed here remain qualitatively valid in case of the benchmarks BP3a, BP3b, BP4a and BP4b as can be read from Fig.~\ref{relic3to4}. The correction fraction for $\l_L = 0.002$ is however smaller compared to the $\l_L = 0.001$ case. This does not come as a surprise since the tree level magnitude is less in BP1a and BP1b. The corresponding corrections for BP3a (BP3b) stand at around $-16.25 \; (-10.64)\%, \; -23.45 \; (-18.58)\%$ and $-46.04 \; (-43.15)\%$ respectively for $\l_2 = 0.1, 1$ and 5.

\begin{figure} 
\begin{center}
\includegraphics[scale=0.38]{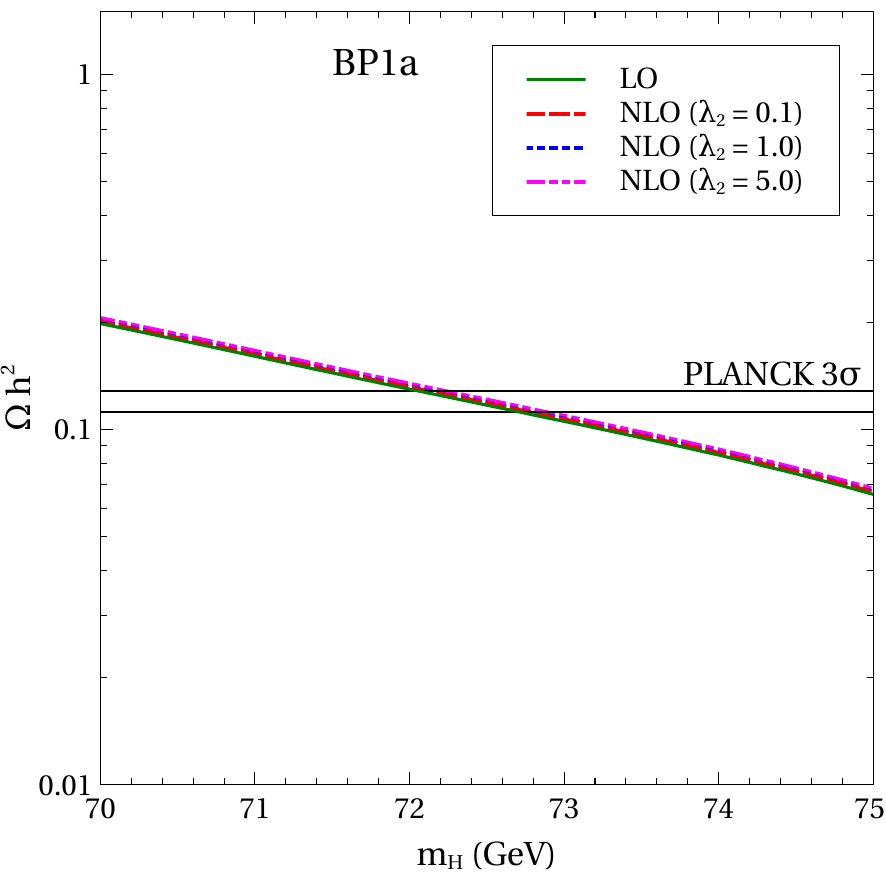}~~~ 
\includegraphics[scale=0.38]{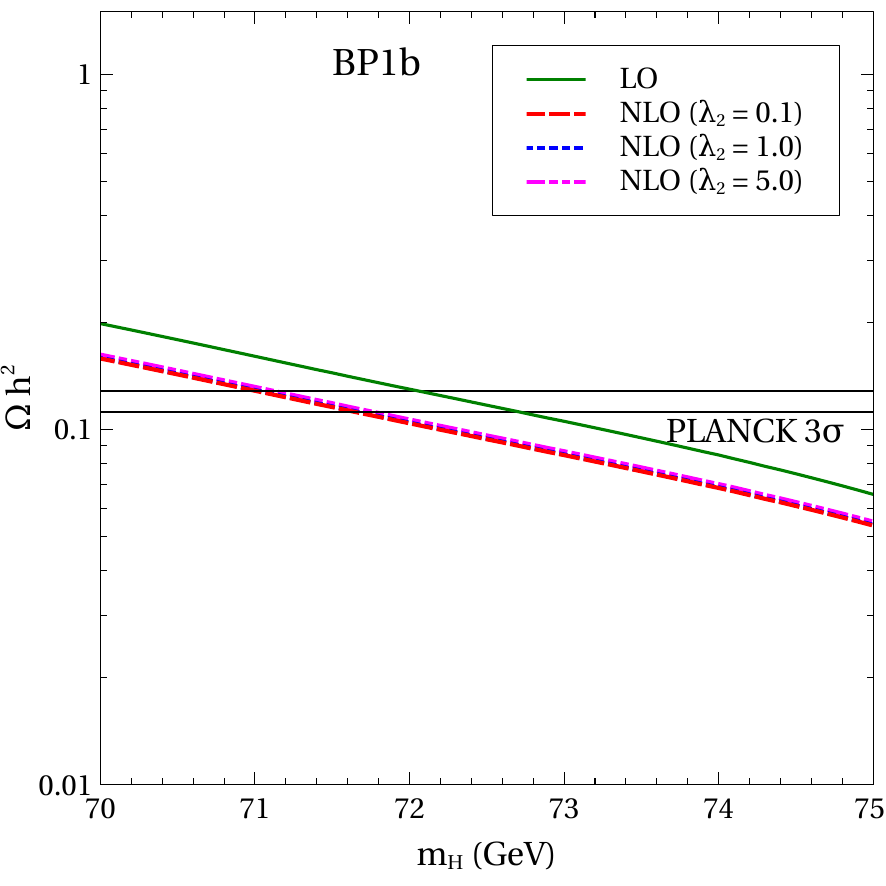}
\\
\includegraphics[scale=0.38]{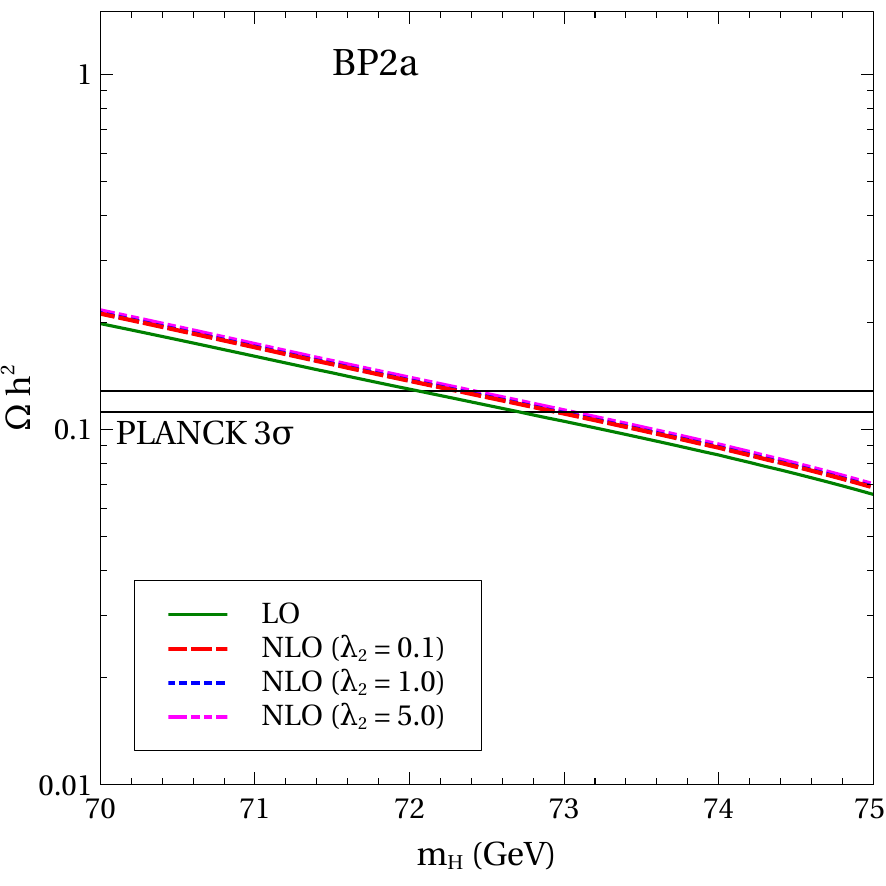}~~~ 
\includegraphics[scale=0.38]{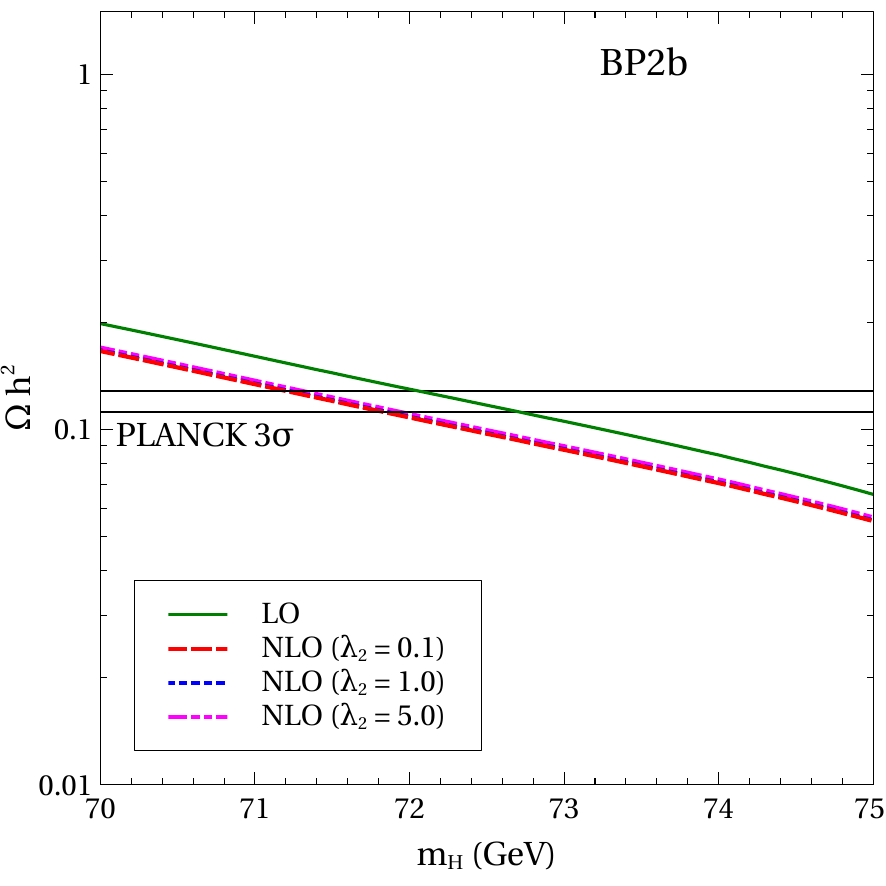}
\caption{Variation of the thermal relic with the DM mass (70 GeV $< m_H <$ 80 GeV). The tree level and one-loop values are denoted by the green solid and red dashed ($\l_2 = 0.1$), blue dotted ($\l_2 = 1$) and pink dot-dashed ($\l=5$) respectively.}
\label{relic1to2_2}
\end{center}
\end{figure}

\begin{figure} 
\begin{center}
\includegraphics[scale=0.38]{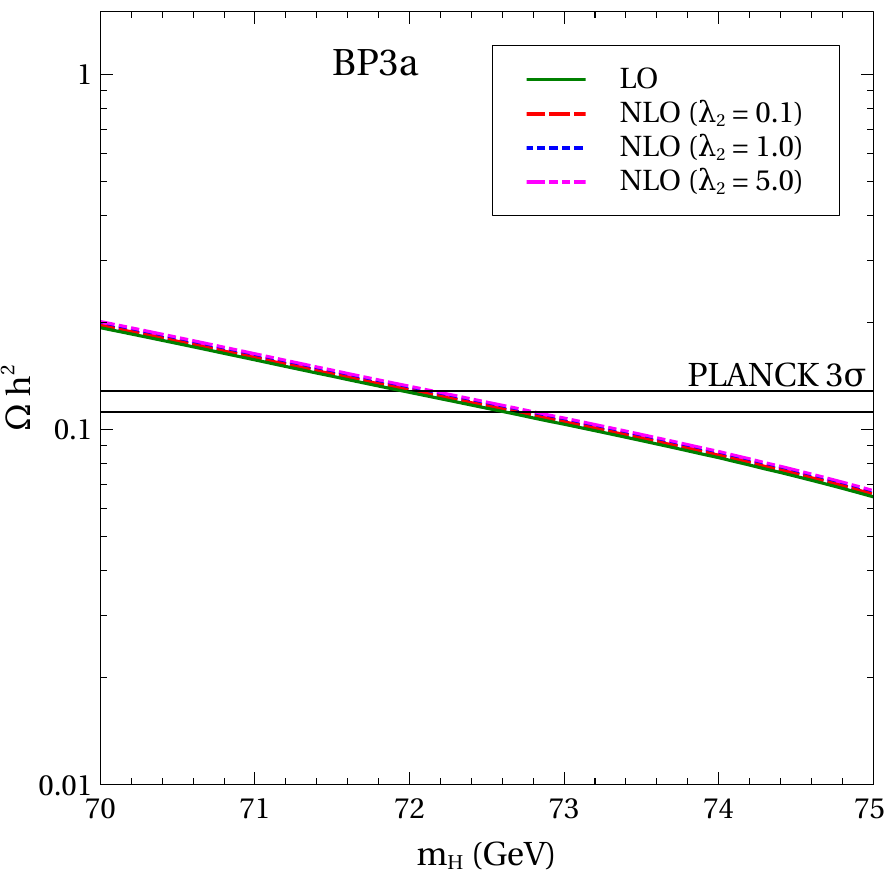}~~~ 
\includegraphics[scale=0.38]{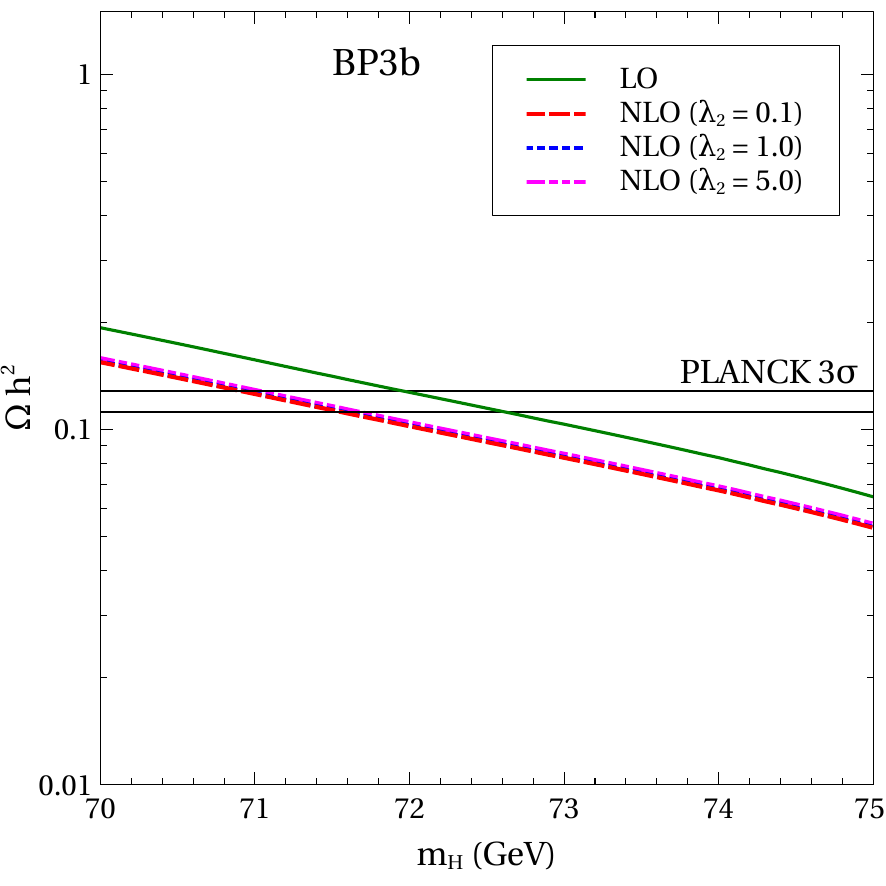}
\\
\includegraphics[scale=0.38]{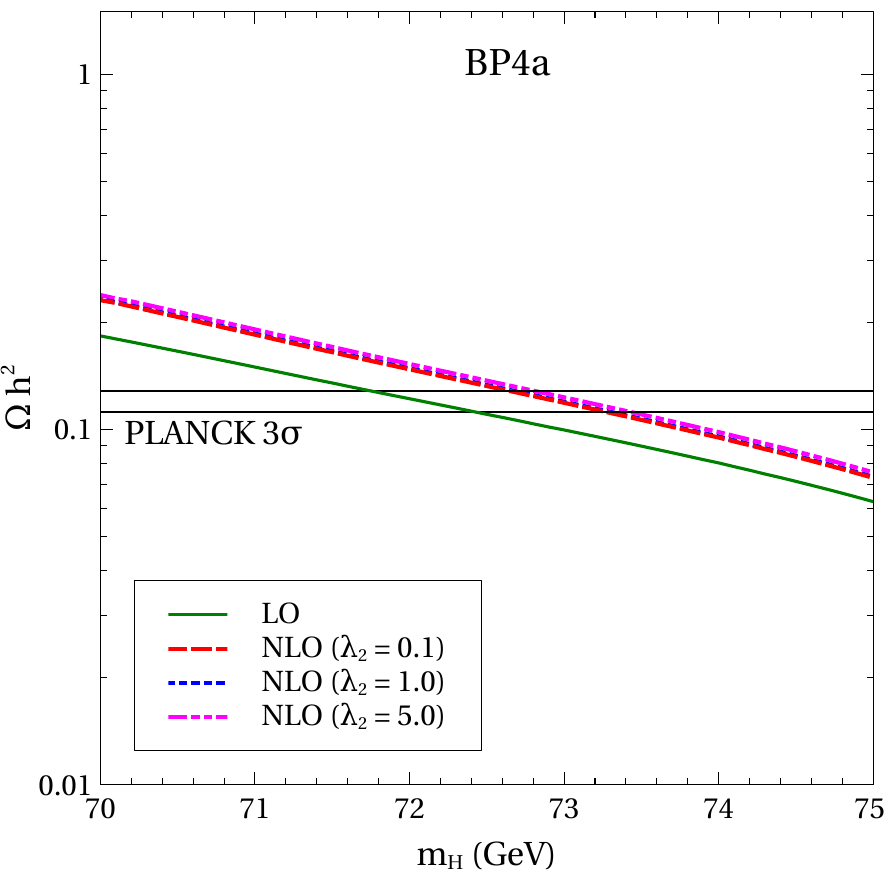}~~~ 
\includegraphics[scale=0.38]{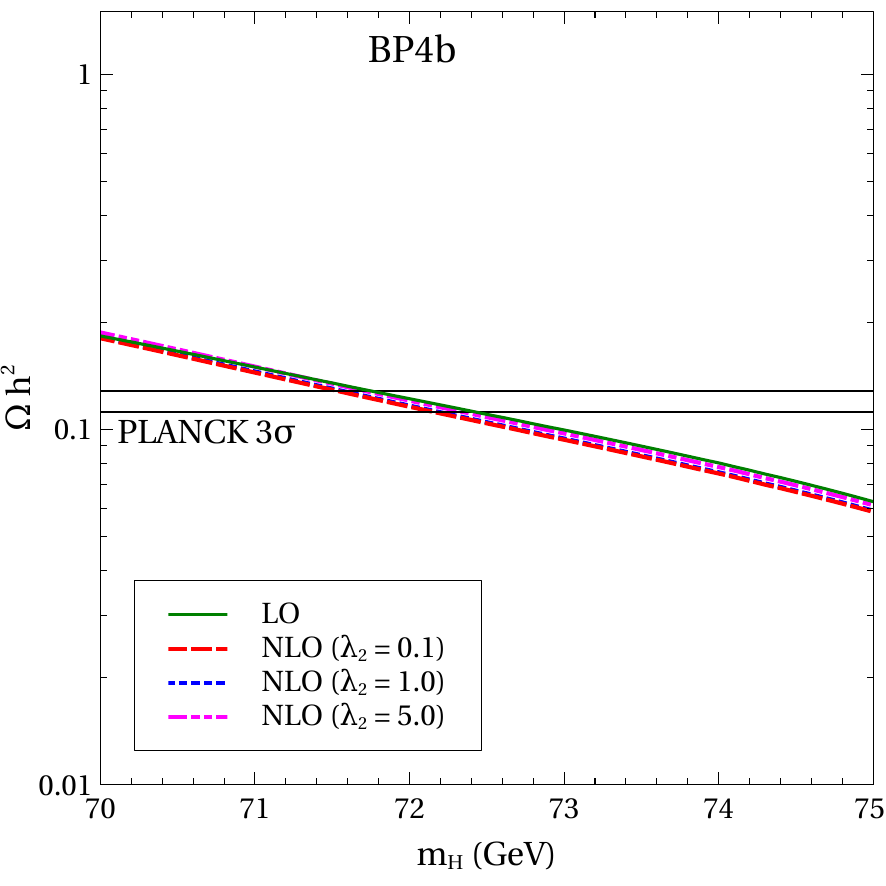}
\caption{The same as in Fig.~\ref{relic1to2_2} for BP3a, BP3b, BP4a and BP4b.}
\label{relic3to4_2}
\end{center}
\end{figure}

The second mass point near the funnel region ($m_H > m_h/2$) features negative value of the 1PI form factor. The relic becomes less sensitive to the portal coupling and the ensuing loop effects since annihilation to 3-body final states (driven by gauge interactions) open up. The shift in the relic is illustrated in Figs.~\ref{relic1to2_2} and~\ref{relic3to4_2}. In fact, it is difficult to discern the effect of a loop-corrected $\Omega h^2$ from the PLANCK 3$\sigma$ uncertainty band, in this region. The only exceptions are BP1b, BP3b and BP4a. Therefore, we conclude that the mass point in the $m_H < m_h/2$ region is most susceptible to a loop-corrected $H-H-h$ interaction. 

\begin{figure} 
\begin{center}
\includegraphics[scale=0.38]{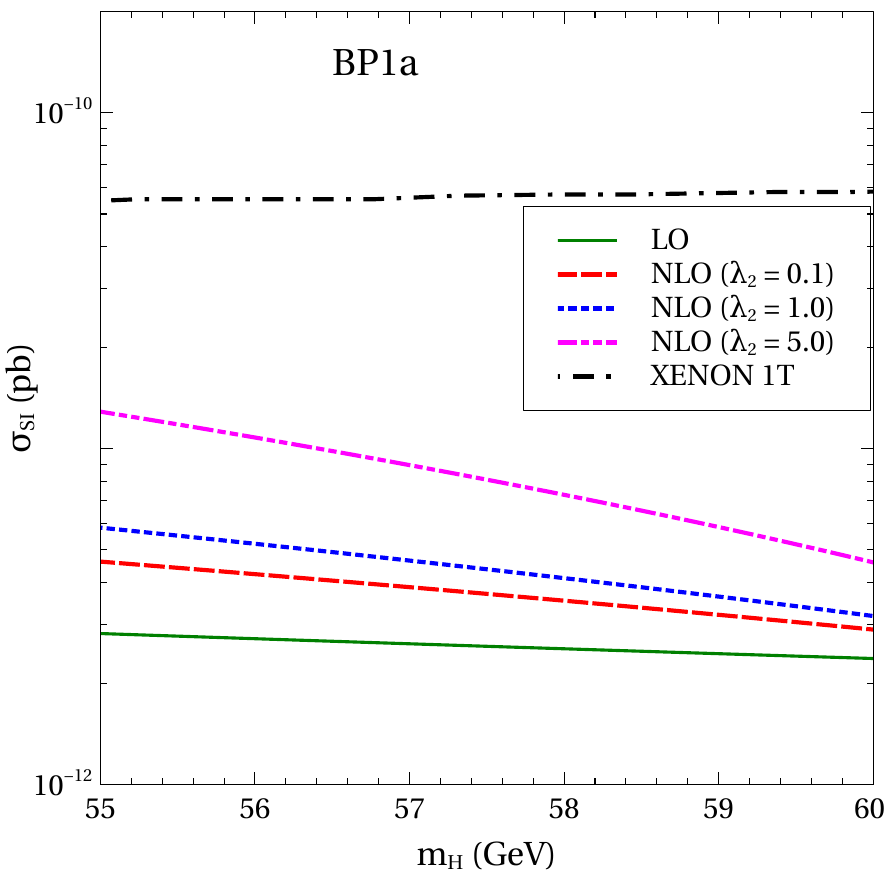}~~~ 
\includegraphics[scale=0.38]{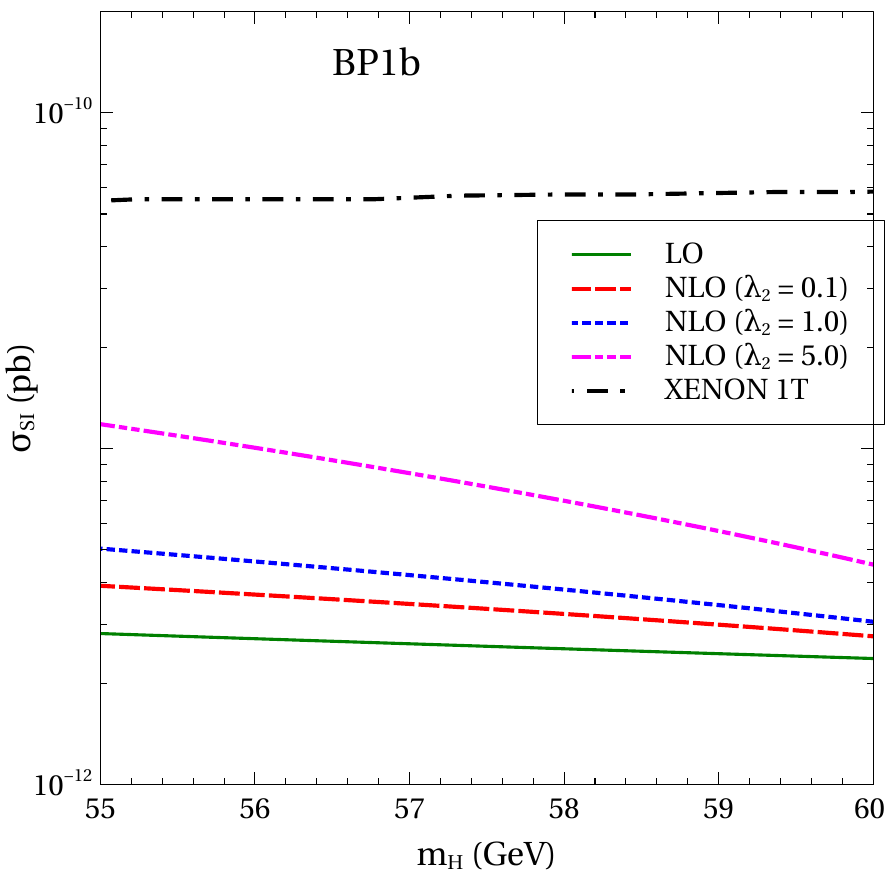}
\includegraphics[scale=0.38]{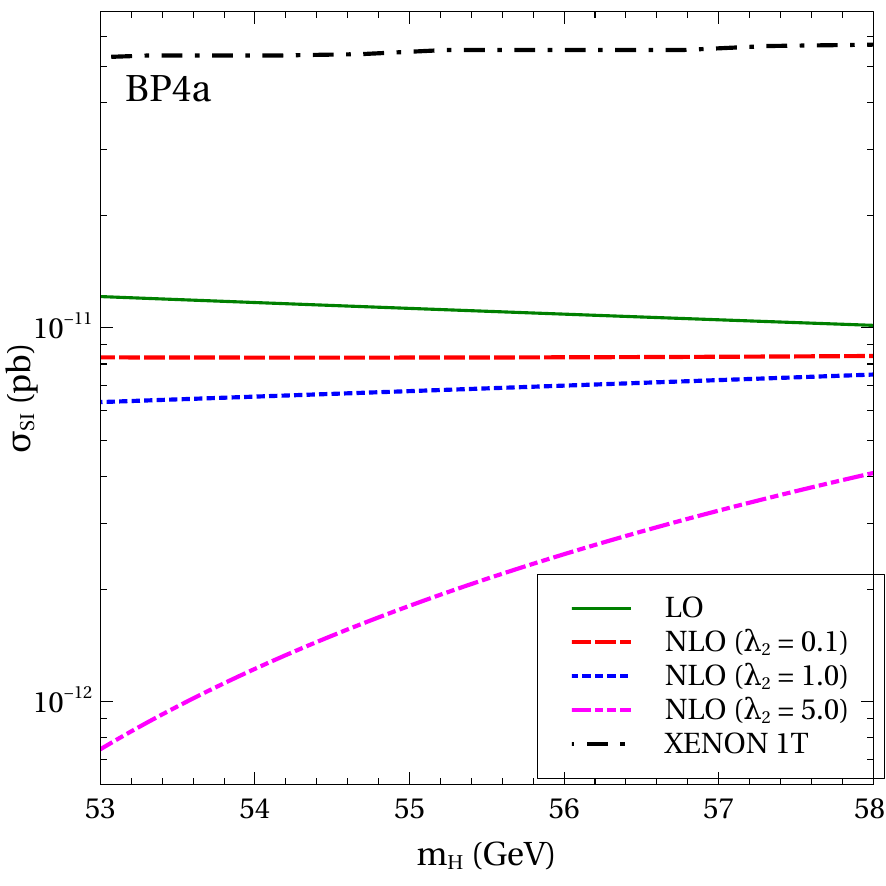}~~~ 
\includegraphics[scale=0.38]{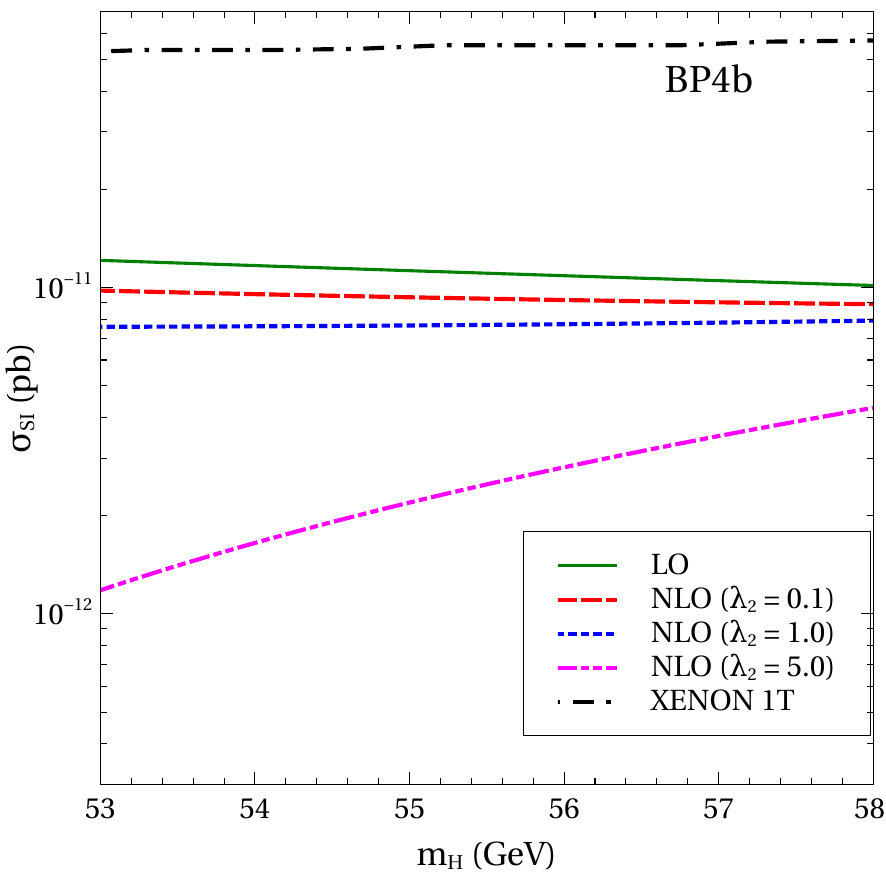}
\caption{Variation of the spin-independent cross section with the DM mass for BP1a, BP1b, BP4a and BP4b. The color coding can be read from the legends.}
\label{DD}
\end{center}
\end{figure}

Fig.~\ref{DD} shows the change in the spin-independent dark matter-nucleon scattering rate under radiative corrections\footnote{Ref. \cite{Klasen:2013btp} examines one-loop corrections to the direct detection process. However, they seem to take into account only the amplitudes involving the gauge couplings. This is somewhat complimentarity to our approach where we focus on correcting the scalar portal coupling, and, this does include the gauge bosons running in the loops. Near $m_H = 57$ GeV for BP1a, the direct detection cross section increases by a factor of $\sim$ 4 when $\l_2$ = 5 is taken. And this is comparable to a $\sim$ 2-3 fold enhancement reported in the aforementioned study. Both these numbers  surely lead us to the correct ballpark nonetheless. That said, the most accurate figure will only emerge when the calculation is not restricted to a particular set of diagrams.}. The rate being proportional to $|\l_L v$ + $\Gamma^{\text{ren}}_{H H h}( m^2_H,m^2_H,4m^2_H)|^2$, it increases upon increasing the effective portal coupling through loop effects and vice versa. An understanding of the aforementioned features governing the strength of the renormalised one-loop form factor therefore suffices to predict the loop-corrected direct detection rates. We plot the direct detection rates for BP1a, BP1b, BP4a and BP4b in Fig.~\ref{DD} near the $m_H < m_h/2$ mass point. One naturally witnesses a higher direct detection cross section upon incorporating loop corrections in case of BP1a and BP1b. However, it still stays within the XENON 1T bound. On the other hand, the latter two benchmarks are characterised by $\l_L < 0$ and hence this opens up the possibility of the cancellation between the tree level and the 1PI amplitudes. The cancellation is obviously maximum for $\l_2 = 5$ and therefore the corresponding curves leads to the lowest direct detection rates as can be read from Fig.~\ref{DD}. Such plots for the other benchmark points are not shown for brevity. 

\begin{figure} 
\begin{center}
\includegraphics[scale=0.38]{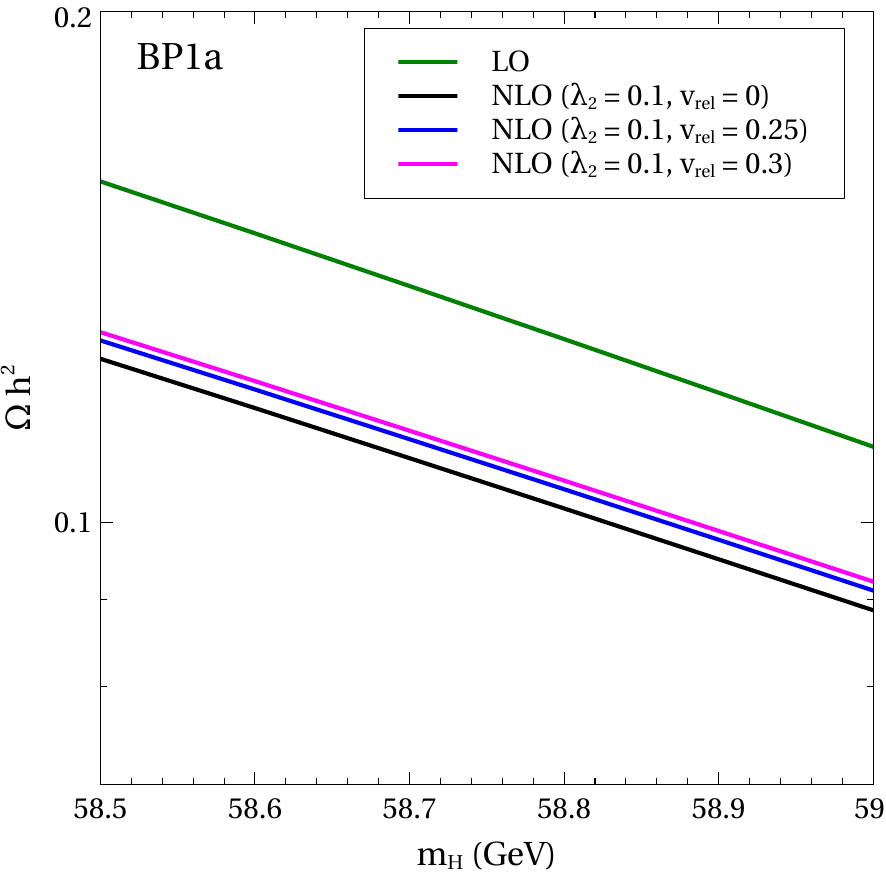}~~~ 
\caption{Effects of $v_{\text{rel}} \neq 0$.}
\label{relic_v_nonzero}
\end{center}
\end{figure}

The aforementioned discussion applies to the case where the DM annihilates with an exactly zero relative velocity. This is a reasonable approximation to the actual scenario where such non-relativistic annihilations are indeed at play. However, for the sake of completeness, it is useful to demonstrate the effect of a non-zero velocity on the NLO relic. Noting that $v_{\text{rel}} \simeq \sqrt{\frac{2}{x_F}}$ and typically $x_F \sim 20 - 30$, one obtains $v_{\text{rel}} \sim 0.25 - 0.3$. We therefore use $p^2 = 4 m^2_H (1 + \frac{v_{\text{rel}}^2}{4}) + \mathcal{O}(v_{\text{rel}}^4)$ with $v_{\text{rel}} = 0.25,~0.3$ and choose BP1a to demonstrate the ensuing effect. The corresponding relic curves are shown in Fig.~\ref{relic_v_nonzero}.

The NLO relics at $m_H = 58.7$ for $v_{\text{rel}} = 0, ~0.25, ~0.3$ are 0.109, 0.112 and 0.113 respectively. This implies that the percentage change noted in going from $v_{\text{rel}}$ = 0 to $v_{\text{rel}}$ = 0.3 is $\simeq$ 3.67$\%$. This is meagre compared to the $\sim$ 25$\%$ correction obtained in incorporating a zero velocity NLO contribution to the LO at the same $m_H$. A similar behaviour is seen or qualitatively expected upon changing $m_H$ or the BP itself. One hence concludes that using of $v_{\text{rel}} \neq 0$ in our calculation only leads to a subleading change in the thermal relic. It is thus possible to disentangle the effect of DM self-interaction on the relic density regardless of the typical value of the freeze-out velocity. Our NLO results for $v_{\text{rel}} = 0$ therefore stand as realistic estimates of the strength of radiative corrections to the $H-H-h$ portal coupling.

\subsection{$m_H > 500.0$ GeV}

\begin{figure}[!htb]
\begin{center}
\includegraphics[scale=0.38]{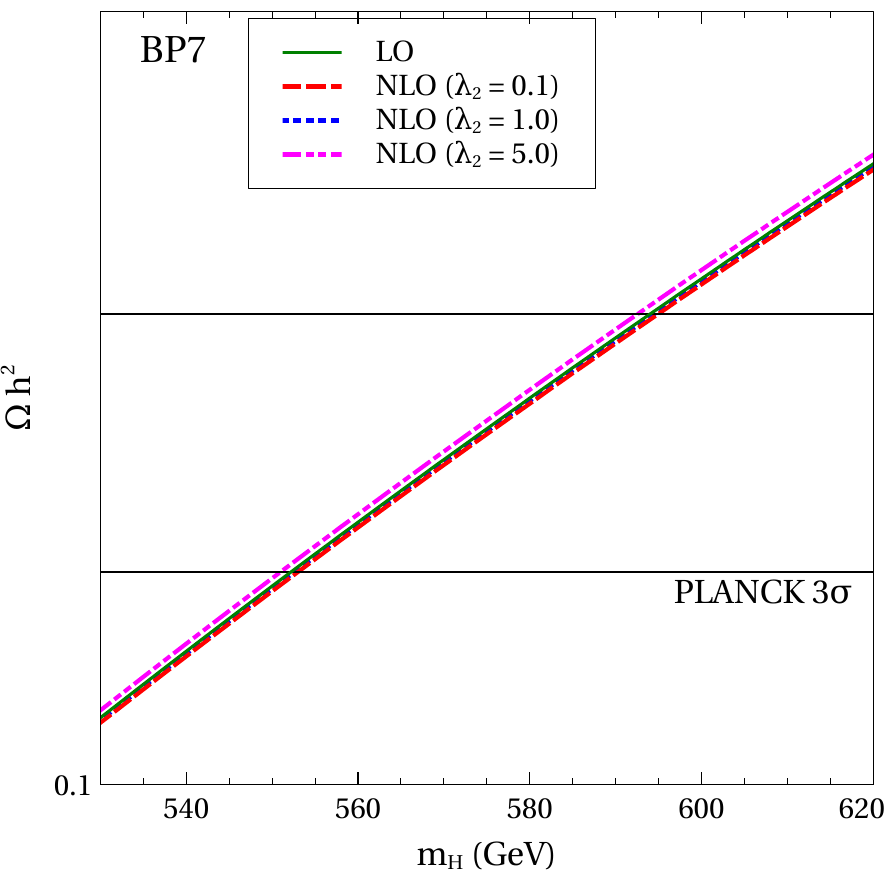}~~~ 
\includegraphics[scale=0.38]{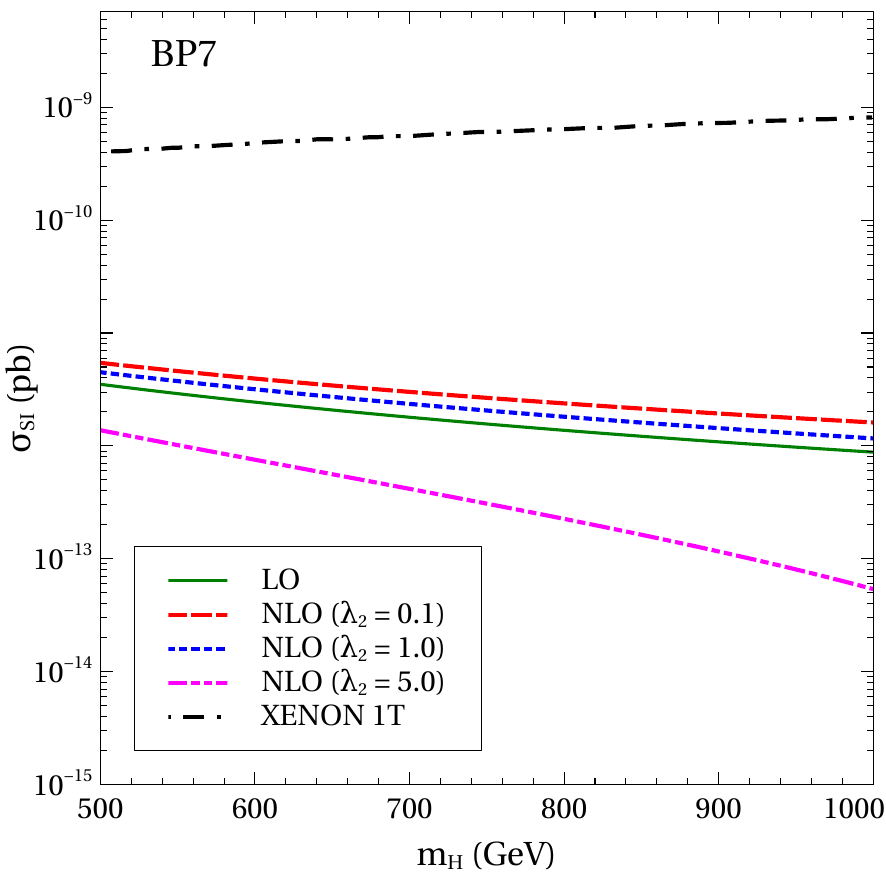}
\caption{Variation of the relic density (left) and the spin-independent cross section (right) with the DM mass. The tree level and one-loop values are denoted by the blue and red curves respectively.}
\label{BP7}
\end{center}
\end{figure}

For this part, we choose $m_A = m_H + 1$ GeV and $m_{H^+} = m_H + 2$ GeV to trigger the requisite co-annihilations, along with setting $\l_L = 0.01$. The radiative correction to $\l_L$ is found to be either positive or negative depending on the value of $\l_2$ (see Fig.~\ref{BP7}). It is approximately $-42.96\%$ for $\l_2 = 5$ and $26.67 \%$ for $\l_2 = 0.1$ near $m_H = 580$ GeV. Though these numbers look sizable, the relic density changes only slightly \emph{w.r.t.} its tree level value. In fact, such a small change is almost indistinguishable from the PLANCK error band and this result remains qualitatively similar for other parameter points. This result does not come as a surprise because the dominant fraction of the relic in this mass region is generated by annihilation and co-annihilation processes of the type $\phi \phi \longrightarrow V V$ where gauge interactions are at play. A more complete picture of loop corrections to the thermal relic is therefore expected to emerge only after loop corrections are incorporated in the gauge interactions. In case of direct detection, the corresponding amplitude features the $H-H-h$ coupling and therefore, loop correction to this coupling is expected to match a full radiative correction to a reasonable degree. It is seen that the deviations in relic density and direct detection rates, \emph{w.r.t.} the tree level, increase as one considers higher values of the inert scalar masses, even if the mass-splitting is kept fixed. This is due to the fact that the parameters $\l_3, \; \l_4$ and $\l_5$ can individually grow in magnitude in the process, thereby increasing the appropriate one-loop form factors.

We remind the readers that one should also include one-loop corrections to the co-annihilation processes for a more accurate analysis. The crucial co-annihilation mediating $HZA$, $HW^{\mp}H^{\pm}$ and $H^{\pm}H^{\mp}Z/\gamma$ vertices shall also receive potentially large corrections from the extended Higgs sector. In addition, a complete NLO analysis of the DM-nucleon scattering rates requires going beyond computing loop corrections to the portal coupling only. To give an example, one-loop triangle graphs with $A$ and $Z$ in the internal lines will be encountered. It is also customary to examine the dependence of the results on the renormalisation scheme chosen. A more exhaustive radiative treatment of the present scenario is currently under preparation. 

\section{Summary and outlook}
\label{sec6}

In this study, we have evaluated one-loop radiative corrections to the dark matter-Higgs portal interaction in the context of the inert doublet model (IDM). Canonical constraints from vacuum stability, perturbative unitarity and LHC data have been taken into account. The motivation behind this work was to obtain a measure of deviation from the leading order results, given that an additional doublet furnishes more bosonic degrees of freedom that can participate in a next-to-leading order analysis. The present renormalisation scheme is based on demanding an unchanged $h \rightarrow H H$ decay width, upon adding the 1PI amplitudes and the counterterms. We have restricted our numerical analysis to a set of representative and somewhat conservative benchmark points that encompass the salient features.

For the dark matter lighter than $m_W$, the inert doublet cannot be fully decoupled even if the other inert scalars are taken to be heavy. This non-decoupling effect induces sizeable loop corrections in the observable quantities and this effect is more prominent for $m_H < m_h/2$. In this region, the radiatively corrected interaction can grow or diminish in magnitude depending on whether the tree level coupling respectively carries a positive or negative sign. On the other hand, the $m_H > 500$ GeV region witnesses comparatively small radiative shifts to the relic abundance. This can be safely attributed to the fact that the portal interaction plays only a subdominant role in this mass region. And therefore, a complete picture can only emerge if one incorporates one-loop effects to all the relevant interactions. Moreover, the radiative corrections will also affect the search prospects of a dark matter particle at the colliders, a promising channel to probe being the monojet + $\slashed{E}_T$ final state. A more exhaustive study on the impact of one-loop corrections to all the relevant interactions in this scenario is presently underway. In all, scalar dark matter models have served as popular frameworks to study interactions of dark matter with the visible world. Through the present study, we have tried to highlight the fact that these models can be made more accurate once they are augmented with relevant radiative corrections. 

\section*{Acknowledgements}

We thank Sun Hao for his help on \texttt{Looptools}. We are grateful to Fawzi Boudjema for his suggestions during the preparation of the manuscript. We also thank Mikael Chala and Andreas Goudelis for providing important feedback to the manuscript. SB acknowledges the support of the Indo French LIA THEP (Theoretical high Energy Physics) of the CNRS and a Durham Junior Research Fellowship COFUNDed between Durham University and the European Union under grant agreement number 609412. NC acknowledges the funding available from the Department of Atomic Energy, Government of India, for the Regional Centre for Accelerator based Particle Physics (RECAPP), Harish-Chandra Research Institute.

\newpage

\appendix

\section{Two-point amplitudes}
\label{appA}
We list all the relevant 1PI amplitudes here. 
\footnotesize
\bea
\Pi^{\text{1PI}}_{h h} (p^2) &=& \frac{1}{16 \pi^2}\Bigg[\frac{3 \l_1}{2} A_0(m^2_h) + \l_1 A_0(m^2_{G^+}) + \frac{\l_1}{2} A_0(m^2_A) + \frac{ (\l_3 + \l_4 + \l_5)}{2} A_0(m^2_H)  + \frac{ (\l_3 + \l_4 - \l_5)}{2} A_0(m^2_A)\nonumber \\
&&
 +\l_3 A_0(m^2_{H^+}) + \frac{9 \l^2_1}{2} v^2 B_0(p^2,m^2_h,m^2_h) + \frac{(\l_3 + \l_4 + \l_5)^2}{2}  v^2 B_0(p^2,m^2_H,m^2_H)\nonumber \\
 &&
 + \frac{\l^2_1}{2}  v^2 B_0(p^2,m^2_{G^0},m^2_{G^0}) + \l^2_1 v^2 B_0(p^2,m^2_{G^+},m^2_{G^+}) + \frac{(\l_3 + \l_4 - \l_5)^2}{2}  v^2 B_0(p^2,m^2_A,m^2_A) \nonumber \\
 &&
 + \l^2_3 v^2 B_0(p^2,m^2_{H^+},m^2_{H^+}) - 2 N_c y^2_t [2 A_0(m^2_t) + (4 m^2_t - p^2)B_0(p^2,m^2_t,m^2_t)] + \frac{d}{2}g^2 A_0(m_W^2) \nonumber \\
 &&
 + \frac{d}{4}(g^2 + {g^{\prime}}^2) A_0(m_Z^2) + \frac{d}{4} v^2 g^4 B_0(p^2,m^2_W,m^2_W)
  + \frac{d}{8} v^2 (g^2 + {g^{\prime}}^2)^2 B_0(p^2,m^2_Z,m^2_Z) \nonumber \\
 &&
 - \frac{g^2}{2}[2 A_0(m_W^2) - A_0(m^2_{G^+}) + (2 p^2 + 2 m^2_{G^+} - m^2_W)B_0(p^2,m^2_W,m^2_{G^+})] \nonumber \\
 && 
 - \frac{g^2 + {g^{\prime}}^2}{4}[2 A_0(m_Z^2) - A_0(m^2_{G^0}) + (2 p^2 + 2 m^2_{G^0} - m^2_Z)B_0(p^2,m^2_Z,m^2_{G^0})]\Bigg]  \\ \nonumber \\ 
\Pi^{\text{1PI}}_{H H} (p^2) &=&  \frac{1}{16 \pi^2}\Bigg[ \frac{(\l_3 + \l_4 + \l_5)}{2} A_0(m^2_h) + \l_3 A_0(m^2_{G^+}) + \frac{(\l_3 + \l_4 - \l_5)}{2} A_0(m^2_{G^0}) + \frac{3\l_2}{2}A_0(m^2_{H}) + \l_2 A_0(m^2_{H^+}) \nonumber \\
 &&
+ \frac{\l_2}{2}A_0(m^2_{A})
+ (\l_3 + \l_4 + \l_5)^2 v^2 B_0(p^2,m^2_h,m^2_H) + \l_5^2 v^2 B_0(p^2,m^2_A,m^2_{G^0}) \nonumber \\
&&
 + \frac{(\l_4 + \l_5)^2}{2} v^2 B_0(p^2,m^2_{H^+},m^2_{G^+}) + \frac{d}{2} g^2 A_0(m_W^2) + \frac{d}{4}(g^2 + {g^{\prime}}^2) A_0(m_Z^2) \nonumber \\
 && 
- \frac{g^2}{2}[2 A_0(m_W^2) - A_0(m^2_{H^+}) + (2 p^2 + 2 m^2_{H^+} - m^2_W)B_0(p^2,m^2_W,m^2_{H^+})] \nonumber \\
 && 
 - \frac{g^2 + {g^{\prime}}^2}{4}[2 A_0(m_Z^2) - A_0(m^2_{A}) + (2 p^2 + 2 m^2_{A} - m^2_Z)B_0(p^2,m^2_Z,m^2_A)]\Bigg] \nonumber \\ \\
\Pi^{\text{1PI}}_{W W,T} (p^2) &=& \frac{1}{16 \pi^2}\Bigg[\frac{g^2}{4}\left[B_5 \left(p^2 ; m_H , m_{H^\pm}\right) + 
 B_5 \left(p^2 ; m_A ,  m_{H^\pm}\right)\right] + g^2 \left( m_W^2 B_0 + \frac 1 4 B_5\right) \left(p^2 ; m_h , m_W\right) \nonumber \\
 && + g^2 \left[\left(\frac 1 4 + 2 c_W^2\right) B_5 + \left(m_W^2 - 
 4 s_W^2 m_W^2 + m_Z^2 - 8 p^2 c_W^2\right) B_0\right]\left(p^2 ; m_Z , m_W\right) \nonumber\\
 && + 2 s_W^2 \left[ B_5 + \left(2 m_W^2 - 4 p^2\right) B_0\right]\left(p^2 ; m_\gamma , m_W\right) - \frac 2 3 g^2 p^2\Bigg]\,
\eea

The definition of the $B_5(p^2,m_1^2,m_2^2)$ function can be found in ~\cite{Hagiwara:1994pw}.

\section{Three-point amplitudes}
\label{appB}
\bea
\Gamma^{\text{1PI}}_{H H h}(p_1^2,p_2^2,q^2) &=& \frac{1}{16 \pi^2}\Bigg[\frac{3}{2}\l_1 (\l_3 + \l_4 + \l_5) v B_0(q^2,m^2_h,m^2_h) + (\l_3 + \l_4 + \l_5)^2 v B_0(p_1^2,m^2_h,m^2_H)\nonumber \\
& &
+ (\l_3 + \l_4 + \l_5)^2 v B_0(p_2^2,m^2_h,m^2_H) + 3 \l_1 (\l_3 + \l_4 + \l_5)^2 v^3 C_0(q^2,p^2_1,p^2_2,m^2_h,m^2_H,m^2_H)\nonumber \\
& &
+ (\l_3 + \l_4 + \l_5)^3 v^3 C_0(q^2,p^2_1,p^2_2,m^2_h,m^2_H,m^2_H)+ \frac{3}{2}\l_2 (\l_3 + \l_4 + \l_5) v B_0(q^2,m^2_H,m^2_H)\nonumber \\
& &
 + \l_1 \l_3 v B_0(q^2,m^2_{G^+},m^2_{G^+}) + \frac{1}{2}\l_1 (\l_3+ \l_4 - \l_5) v B_0(q^2,m^2_{G^0},m^2_{G^0}) +
 \l_5^2 v B_0(p_1^2,m^2_{G^0},m^2_{A})\nonumber \\
& &
  + \l_5^2 v B_0(p_2^2,m^2_{G^0},m^2_{A}) + \frac{1}{2}\l_2 (\l_3+ \l_4 - \l_5) v B_0(q^2,m^2_{A},m^2_{A})\nonumber \\
& &
+ 4 \l_1^2 \l_5 v^3 C_0(q^2,p_1^2,p_2^2,m^2_A,m^2_{G^0},m^2_{G^0}) + \l_5^2(\l_3 + \l_4 - \l_5)v^3 C_0(q^2,p_1^2,p_2^2,m^2_{G^0},m^2_A,m^2_A)\nonumber \\
& &
+ \l_2 \l_3 v B_0(q^2,m^2_{H^+},m^2_{H^+}) + \frac{1}{2}(\l_4 + \l_5)^2 v B_0(p_1^2,m^2_{G^+},m^2_{H^+}) + \frac{1}{2}(\l_4 + \l_5)^2 v B_0(p_2^2,m^2_{G^+},m^2_{H^+})\nonumber \\
& &
+ \l_1(\l_4 + \l_5)^2 v^3 C_0(q^2,p_1^2,p_2^2,m^2_{H^+},m^2_{G^+},m^2_{G^+}) + \l_3(\l_4 + \l_5)^2 v^3 C_0(q^2,p_1^2,p_2^2,m^2_{G^+},m^2_{H^+},m^2_{H^+}) ]\nonumber \\
&& 
+ \frac{d}{4}g^4 v B_0(q^2,m^2_W,m^2_W) + \frac{d}{8}(g^2 + {g^{\prime}}^2)^2 v B_0(q^2,m^2_Z,m^2_Z)\nonumber \\
&&
- \frac{(\l_3 + \l_4 - \l_5)}{4} (g^2 + {g^{\prime}}^2) \Big[p_1^2 C_{21} + p_2^2 C_{22} + 2 p_1.p_2 C_{23} + 4C_{24} + 2p_1.p_2 C_{11} + 2 p_2^2 C_{12}\nonumber \\
&&
 + (-2 p_1.p_2 - p_1^2)C_0\Big](p_1^2,p_2^2,q^2,m^2_{A},m^2_Z,m^2_{A})\nonumber \\
&&
- \frac{\l_3}{2} g^2 \Big[p_1^2 C_{21} + p_2^2 C_{22} + 2 p_1.p_2 C_{23} + 4C_{24} + 2p_1.p_2 C_{11} + 2 p_2^2 C_{12}\nonumber \\
&&
 + (-2 p_1.p_2 - p_1^2)C_0\Big](p_1^2,p_2^2,q^2,m^2_{H^+},m^2_W,m^2_{H^+})\nonumber \\
&&
- \frac{(g^2 + {g^{\prime}}^2)^2}{8} v \Big[p_1^2 C_{21} + p_2^2 C_{22} + 2 p_1.p_2 C_{23} + 4C_{24} +
(3 p_1^2 - p_1.p_2) C_{11} + (3 p_1.p_2 - p_2^2) C_{12}\nonumber \\
&&
 + (2 p_1^2 - 2 p_1.p_2)C_0\Big](p_1^2,p_2^2,q^2,m^2_{Z},m^2_{A},m^2_Z)\nonumber \\
&&
- \frac{g^4}{4} v \Big[p_1^2 C_{21} + p_2^2 C_{22} + 2 p_1.p_2 C_{23} + 4C_{24} +
(3 p_1^2 - p_1.p_2) C_{11} + (3 p_1.p_2 - p_2^2) C_{12}\nonumber \\
&&
 + (2 p_1^2 - 2 p_1.p_2)C_0\Big](p_1^2,p_2^2,q^2,m^2_{W},m^2_{H^+},m^2_W)
\eea

\providecommand{\href}[2]{#2}
\addcontentsline{toc}{section}{References}
\bibliographystyle{JHEP}
\bibliography{refs} 

\end{document}